\documentclass[preprint,aps,11pt,showpacs,nofootinbib,tightenlines]{revtex4}

\usepackage[section]{placeins}
\usepackage{graphicx}
\usepackage{dcolumn}
\usepackage{bm}

\usepackage{epsfig}
\usepackage{amssymb}
\usepackage{amsmath}
\usepackage{flafter}
\usepackage{array}

\usepackage[dvips,usenames]{color}

\newcommand{\slsh}{\!\!\!\!\!/\,}

\newlength{\dinwidth}
\newlength{\dinmargin}
\begin{document}
\title{QCD Factorization Based on Six-Quark Operator Effective Hamiltonian
from Perturbative QCD and Charmless Bottom Meson Decays $B_{(s)}\to
\pi\pi,\pi K, KK$}
\author{Fang Su$^{\ast \dagger}$, Yue-Liang Wu$^{\ast}$, Yi-Bo Yang$^{\ast\ddagger}$ and Ci Zhuang$^{\ast}$ }
\affiliation{$^\ast$ Kavli Institute for Theoretical Physics China,
Institute of Theoretical Physics \\ Chinese Academy of Science,
Beijing 100190, China \\
$^\dagger$ Institut f\"ur Theoretische Physik E, RWTH Aachen,
D-52056, Aachen, Germany \\
$^\ddagger$ Graduate School of the Chinese Academy of Sciences,
Beijing, 100039, China}

\begin{abstract}
\noindent The charmless bottom meson decays are systematically
investigated based on an approximate six quark operator effective
Hamiltonian from perturbative QCD. It is shown that within this
framework the naive QCD factorization method provides a simple way
to evaluate the hadronic matrix elements of two body mesonic decays.
The singularities caused by on mass-shell quark propagator and gluon
exchanging interaction are appropriately treated. Such a simple
framework allows us to make theoretical predictions for the decay
amplitudes with reasonable input parameters. The resulting
theoretical predictions for all the branching ratios and $CP$
asymmetries in the charmless $B^0,\ B^+,\ B_s\to \pi\pi,\ \pi K,\
KK$ decays are found to be consistent with the current experimental
data except for a few decay modes. The observed large branching
ratio in $B\to \pi^0\pi^0$ decay remains a puzzle though the
predicted branching ratio may be significantly improved by
considering the large vertex corrections in the effective Wilson
coefficients. More precise measurements of charmless bottom meson
decays, especially on CP-violations in $B\to K K$ and $B_s\to
\pi\pi, \pi K, KK$ decay modes, will provide a useful test and guide
us to a better understanding on perturbative and nonperturbative
QCD.
\end{abstract}
\pacs{13.25.Hw,12.38.Bx,12.38.Lg,11.30.Er}

\maketitle

\section{Introduction}

Hadronic B-meson decays play importance role not only for
understanding the dynamical scheme of hadronic decays and testing
the flavor structure of the Standard Model(SM), but also for probing
the origin of CP violation and new physics signals beyond the SM. In
particular, the precise measurement and systematic study for
hadronic charmless B decays may provide a window for such purposes.
The branching ratios of $B \rightarrow \pi \pi$ and $\pi K$ modes
have been measured with a good accuracy\cite{HFAG} and a large
direct CP violation has been established in $\pi^+ K^-$ mode
\cite{HFAG}. The most severe discrepancies between the experimental
data and theoretical predictions come from the unexpected large
branch ratio of $B \to \pi^0\pi^0$ and some unclear CP violations in
$B\to \pi^0 K$ decays, which are called $\pi\pi, \pi K$
puzzles\cite{pikpuzzle,WZ}. Theoretically, to predict consistently
those decays, it needs to deal with the short-distance contributions
in a complete and systematic way from the high energy scale to a
proper low energy scale at which the perturbative calculations
remain reliable, and treat the long-distance contributions which
contain the non-perturbative strong interactions involved in those
decays. The main task is to reliably compute the hadronic matrix
elements between the initial and final hadron states. Several novel
methods based on the naive factorization approach (FA) and four
quark operator effective Hamiltonian have been developed to evaluate
the hadronic matrix elements, such as the QCD factorization approach
(QCDF)\cite{Beneke:1999br}, the perturbation QCD method (pQCD)
\cite{Keum:2000ph}, and the soft-collinear effective theory
(SCET)\cite{Bauer:2000ew}. These methods have been widely used in
analyzing hadronic B-meson decays and made great progresses in
understanding the hadronic structure and properties of strong
interactions. To understand the puzzles whether they are due to the
unknown new physics or it is because of the lack of our knowledge on
the hadronic properties of strong interactions, it still needs to
investigate further the various approaches within the framework of
QCD and to check the validity of assumptions and approximations made
in the practical calculations.

The widely used theoretical framework of weak decays is based on the
current-current four fermion operator effective Hamiltonian derived
via operator product expansion and renormalization group evolution.
In hadronic weak decays, the short-distance contributions of QCD are
characterized by the Wilson coefficient functions of four quark
operators and the long-distance contributions are in principle
obtained by evaluating the hadronic matrix elements of four quark
operators. The Wilson coefficient functions are in general
calculated by perturbative QCD which is well developed, while the
evaluation of hadronic matrix elements remains a hard task as it
involves non-perturbative effects of QCD. To deepen our insights
into the hadronic decays, we shall first reinvestigate the four
quark operator effective Hamiltonian whether it is always suitable
as a basic framework for all hadronic weak decays. In fact, for the
mesonic two body decays of B meson, it concerns three
quark-antiquark pairs once each meson is regarded as the
quark-antiquark bound state at the quark level structure. This fact
then naturally motivates us to consider six-quark operator effective
Hamiltonian instead of four-quark operator effective Hamiltonian.
Namely, we shall begin with six quark diagrams of weak decays with
both W-boson exchange and gluon exchange, and derive formally the
six-quark operator effective Hamiltonian based on operator product
expansion and renormalization group evolution when including loop
corrections of six quark diagrams. We shall show how this approach
allows us to figure out what are the assumptions and approximations
made in effective four quark operator approach, and how the simple
QCD factorization scheme can reliably be applied to evaluate the
hadronic matrix elements with the six quark operator effective
Hamiltonian. For the infrared singularity caused by the gluon
exchanging interaction when evaluating the hadronic matrix elements
of effective six quark operators, it is shown to be simply treated
by the introduction of a mass scale motivated from the gauge
invariant loop regularization method~\cite{LRC}, where the energy
scale $\mu_g$ is introduced to play the role of infrared cut-off
energy scale without violating gauge invariance.

The paper is organized as follows. In section \ref{sec:sqeh}, after
briefly reviewing the four quark operator effective Hamiltonian, we
begin with the primary six quark diagrams with a single W-boson
exchange and a single gluon exchange, and the corresponding initial
six-quark operator. It is shown that a complete six quark operator
effective Hamiltonian is in general necessary to include all
contributions from both perturbative and non-perturbative QCD
corrections, especially the non-pertubative QCD corrections at low
energy scale $\mu < m_c\sim 1.5$ GeV could be sizable. To
demonstrate how the six quark operator effective Hamiltonian
provides a reliable framework for hadronic two body decays of B
meson, we will focus, as a good approximation, on the dominant QCD
loop diagrams of six quarks so as to avoid the tedious calculations.
In section \ref{sec:QCDF}, it is demonstrated how the QCD
factorization approach becomes a simple and natural tool to evaluate
the hadronic matrix elements of mesonic two body decays based on the
six quark operator effective Hamiltonian. In particular, the
so-called factorizable and non-factorizable, emission and
annihilation diagram contributions are automatically the
consequences of QCD factorization for the hadronic matrix elements
of effective six quark operators. The treatment on the singularities
caused by the gluon exchanging interactions and the on mass-shell
fermion propagator is presented in Section \ref{sec:TOD}. In Section
\ref{sec:Amplitude}, all the amplitudes of charmless bottom meson
decays are completely obtained by using the QCD factorization
approach based on the approximate six quark operator effective
Hamiltonian. Our numerical results with appropriate input parameters
are presented in section \ref{sec:nrcpe}, as a good approximation,
the resulting predictions on branching ratios and CP violations of
charmless bottom meson decays are much improved and also more closed
to the current experimental data. The conclusions and remarks are
given in last section. The detailed calculations involved in the
evaluation of various decay amplitudes are presented in the
Appendix.

\section{Effective Hamiltonian of Six Quark Operators}\label{sec:sqeh}

\subsection{Four Quark Operator Effective Hamiltonian}

Let us  start from the four-quark effective operators in the
effective weak Hamiltonian. The initial four quark operator due to
weak interaction via W-boson exchange is given as follows for B
decays
\begin{equation}
O_{1}=(\bar{q}^u_{i}b_{i})_{V-A}(\bar{q}^d_{j}u_{j})_{V-A}, \qquad
q^u=u,\ c, \quad  q^d = d,\ s
\end{equation}
The complete set of four quark operators are obtained from QCD and
QED corrections which contain the gluon exchange diagrams, strong
penguin diagrams and electroweak penguin diagrams. The resulting
effective Hamiltonian(for $b\to s$ transition) with four quark
operators is known to be as follows
\begin{eqnarray}
H_{\rm eff}\, =\, {G_F\over\sqrt{2}} \sum_{q=u,c}
\lambda_q^{s} \left[C_1(\mu)O_1^{(q)}(\mu) +C_2(\mu)O_2^{(q)}(\mu)+
\sum_{i=3}^{10}C_i(\mu)O_i(\mu)\right]+h.c.\;,\label{eq:hpk}
\end{eqnarray}
with $\lambda_q^{s} = V_{qb}V^*_{qs}$ and $V_{ij}$ the CKM matrix
elements, $C_i(\mu)$ the Wilson coefficient functions\cite{4qham}
and $O_i(\mu)$ the four quark operators
\begin{eqnarray}
\begin{array}{ll}
\displaystyle O_1^{(q)}\, =\,
(\bar{q}_ib_i)_{V-A}(\bar{s}_jq_j)_{V-A}\;, & \displaystyle
O_2^{(q)}\, =\,(\bar{s}_ib_i)_{V-A}(\bar{q}_jq_j)_{V-A}\;,
\\
\displaystyle O_3\,
=\,(\bar{s}_ib_i)_{V-A}\sum_{q'}(\bar{q}'_jq'_j)_{V-A}\;,
&\displaystyle O_4\,
=\,\sum_{q'}(\bar{q}'_ib_i)_{V-A}(\bar{s}_jq'_j)_{V-A}\;,
\\
\displaystyle O_5\,
=\,(\bar{s}_ib_i)_{V-A}\sum_{q'}(\bar{q}'_jq'_j)_{V+A}\;,
&\displaystyle O_6\,
=\,-2\sum_{q'}(\bar{q}'_ib_i)_{S-P}(\bar{s}_jq'_j)_{S+P}\;,
\\
\displaystyle O_7\,
=\,\frac{3}{2}(\bar{s}_ib_i)_{V-A}\sum_{q'}e_{q'}(\bar{q}'_jq'_j)_{V+A}\;,&
\displaystyle O_8\, =\,
-3\sum_{q'}e_{q'}(\bar{q}'_ib_i)_{S-P}(\bar{s}_jq'_j)_{S+P}\;,
\\
\displaystyle O_9\,
=\,\frac{3}{2}(\bar{s}_ib_i)_{V-A}\sum_{q'}e_{q'}(\bar{q}'_jq'_j)_{V-A}\;,&
\displaystyle O_{10}\, =\,
\frac{3}{2}\sum_{q'}e_{q'}(\bar{q}'_ib_i)_{V-A}(\bar{s}_jq'_j)_{V-A}\;,\\
\end{array}
\label{eq:o}
\end{eqnarray}
Here the Fermi constant $G_F=1.16639\times
10^{-5}\;{\rm GeV}^{-2}$, and the color indices $i, \ j$,
 and the notations $(\bar{q}'q')_{V\pm A} = \bar q' \gamma_\mu (1\pm
\gamma_5)q'$. The index $q'$ in the summation of the above operators
runs through $u,\;d,\;s$, $c$, and $b$. The effective Hamiltonian
for the $b\to d$ transition
can be obtained by changing $s$ into $d$ in
Eqs.~(\ref{eq:hpk})and (\ref{eq:o}).

\subsection
{\boldmath Six Quark Diagrams and Effective
Operators}\label{sec:sqd}

\begin{figure}[h]
\begin{center}
\includegraphics[scale=0.65]{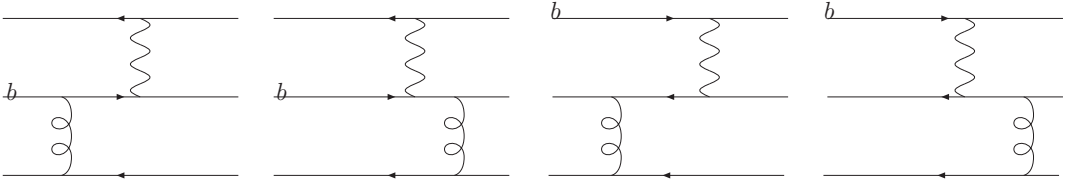}\\
  \caption{Four different six quark diagrams with a single W-boson exchange and a single gluon exchange}\label{pic:4insert}
  \end{center}
\end{figure}

As mesons are regarded as quark and anti-quark bound states, the
mesonic two body decays actually involve three quark-antiquark
pairs. It is then natural to consider the six quark Feynman diagrams
which lead to three effective currents of quark-antiquark. The
initial six quark diagrams of weak decays contain one W-boson
exchange and one gluon exchange, thus there are four different
diagrams as the gluon exchange interaction can occur for each of
four quarks in the W-boson exchange diagram, see Fig.
\ref{pic:4insert}.

The resulting initial effective operators contain four terms
corresponding to the four diagrams, respectively. In a good
approximation, the four quarks via W-boson exchange can be regarded
as a local four quark interaction at the energy scale much below the
W-boson mass, while two QCD vertexes due to gluon exchange are at
the independent space-time points, the resulting effective six quark
operators are hence in general nonlocal. The six-quark operators
corresponding to the four diagrams in Fig. \ref{pic:4insert} are
found to be
\begin{eqnarray}
  O^{(6)}_{q_1}\, &=&\, 4\pi\alpha_s \int\!\!\int  \frac{\emph{d}^4k}{(2\pi)^4}\, \frac{\emph{d}^4p}{(2\pi)^4}\,e^{-i((x_1-x_2)p+(x_2-x_3)k)}
  (\bar{q'}(x_3)\gamma_{\nu}T^{a} q'(x_3))\frac{1}{k^2+i\epsilon}\nonumber\\
  &&(\bar{q}_{2}(x_1)\Gamma_{1}\frac{p\!\!\!/+m_b}{p^2-m_b^2+i\epsilon}\gamma^{\nu}T^{a} q_{1}(x_2))*
  (\bar{q}_{4}(x_1) \Gamma_{2} q_{3}(x_1)),\nonumber\\
  O^{(6)}_{q_2}\, &=&\, 4\pi\alpha_s \int\!\!\int  \frac{\emph{d}^4k}{(2\pi)^4}\, \frac{\emph{d}^4p}{(2\pi)^4}\,e^{-i((x_1-x_2)p+(x_2-x_3)k)}
  (\bar{q'}(x_3)\gamma_{\nu}T^{a} q'(x_3))\frac{1}{k^2+i\epsilon}\nonumber\\
  &&(\bar{q}_{2}(x_2)\frac{p\!\!\!/+m_{q_1}}{p^2-m_{q_1}^2+i\epsilon}\gamma^{\nu}T^{a}\Gamma_{1} q_{1}(x_1))*
  (\bar{q}_{4}(x_1) \Gamma_{2} q_{3}(x_1)),\nonumber\\
  O^{(6)}_{q_3}\, &=&\, 4\pi\alpha_s \int\!\!\int  \frac{\emph{d}^4k}{(2\pi)^4}\, \frac{\emph{d}^4p}{(2\pi)^4}\,e^{-i((x_1-x_2)p+(x_2-x_3)k)}
  (\bar{q'}(x_3)\gamma_{\nu}T^{a} q'(x_3))\frac{1}{k^2+i\epsilon}\nonumber\\
  &&(\bar{q}_{2}(x_1)\Gamma_{1} q_{1}(x_1))*
  (\bar{q}_{4}(x_1) \Gamma_{2} \frac{p\!\!\!/+m_{q_3}}{p^2-m_{q_3}^2+i\epsilon}\gamma^{\nu}T^{a} q_{3}(x_2)),\nonumber\\
  O^{(6)}_{q_4}\, &=&\, 4\pi\alpha_s \int\!\!\int  \frac{\emph{d}^4k}{(2\pi)^4}\, \frac{\emph{d}^4p}{(2\pi)^4}\,e^{-i((x_1-x_2)p+(x_2-x_3)k)}
  (\bar{q'}(x_3)\gamma_{\nu}T^{a} q'(x_3))\frac{1}{k^2+i\epsilon}\nonumber\\
  &&(\bar{q}_{2}(x_1)\Gamma_{1} q_{1}(x_1))*
  (\bar{q}_{4}(x_2)\frac{p\!\!\!/+m_{q_2}}{p^2-m_{q_2}^2+i\epsilon}\gamma^{\nu}T^{a} \Gamma_{2} q_{3}(x_1)),
\label{eq:six}
\end{eqnarray}

where $k$ and $p$ correspond to the momenta of gluon and quark in
their propagators. $q_1$ is usually set to be heavy quark like b
quark. $x_1$, $x_2$ and $x_3$ are space-time points corresponding to
three vertexes. The color index is summed between $q_1,q_2$ and
$q_3,q_4$. Note that all the six quark operators are proportional to
the QCD coupling constant $\alpha_s$ due to gluon exchange. Thus the
initial six quark operator is given by summing over the above four
operators
\begin{eqnarray}
  O^{(6)}=\sum_{j=1}^4 O^{(6)}_{q_j}.
\end{eqnarray}

Actually, the initial six quark operators $O^{(6)}_{q_j}$
($j=1,2,3,4$) can be obtained from the following initial four quark
operator via a single gluon exchange
\begin{eqnarray}
  O \equiv (\bar{q}_{2} \Gamma_{1} q_{1})*(\bar{q}_{4} \Gamma_{2} q_{3}).\label{eq:any}
\end{eqnarray}

\subsection{ Six Quark Operator Effective Hamiltonian via Perturbative QCD}\label{sec:sqehbd}

Based on the above considerations with the introduction of six quark
operators, in this section we shall specify the initial six quark
operator $O_1^{(q)(6)}$ $(q=u,\, c)$ to the case of nonleptonic
bottom hadron decays and show how to obtain six quark operator
effective Hamiltonian. The initial six quark operator in b-decay
with $\Delta S \neq 0$ is as follows (for the $b\to d$ transition
with $\Delta S = 0$, just replacing $s$ by $d$)
\begin{eqnarray}
  O_1^{(q)(6)}\,&=&\sum_{l=1}^4 O^{(q)(6)}_{1q_l} \nonumber\\
  &=&4\pi\alpha_s(m_W)\int\!\!\int  \frac{\emph{d}^4k}{(2\pi)^4}\, \frac{\emph{d}^4p}{(2\pi)}\,e^{-i((x_1-x_2)p+(x_2-x_3)k)}\nonumber\\
  &&\{ \ (\bar{q}_i(x_1)\gamma^{\mu}(1-\gamma^{5})\frac{p\!\!\!/+m_b}{p^2-m_b^2+i\epsilon}\gamma^{\nu}T_{ik}^{a}b_k(x_2))
  (\bar{s}_j(x_1)\gamma_{\mu}(1-\gamma^{5})q_j(x_1))\nonumber\\
  &&+(\bar{q}_k(x_2)\frac{p\!\!\!/+m_q}{p^2-m_q^2+i\epsilon}\gamma^{\nu}T_{ki}^{a}\gamma^{\mu}(1-\gamma^{5})b_i(x_1))
  (\bar{s}_j(x_1)\gamma_{\mu}(1-\gamma^{5})q_j(x_1))\nonumber\\
  &&+(\bar{q}_i(x_1)\gamma^{\mu}(1-\gamma^{5})b_i(x_1))
  (\bar{s}_j(x_1)\gamma_{\mu}(1-\gamma^{5})\frac{p\!\!\!/+m_q}{p^2-m_q^2+i\epsilon}\gamma^{\nu}T_{jk}^{a}q_k(x_2))\nonumber\\
  &&+(\bar{q}_i(x_1)\gamma^{\mu}(1-\gamma^{5})b_i(x_1))
  (\bar{s}_k(x_2)\frac{p\!\!\!/+m_s}{p^2-m_s^2+i\epsilon}\gamma^{\nu}T_{kj}^{a}\gamma_{\mu}(1-\gamma^{5})q_j(x_1)) \ \}\nonumber\\
  &&\frac{1}{k^2+i\epsilon}(\bar{q'}_m(x_3)\gamma_{\nu}T_{mn}^{a} q'_n(x_3)),
  \label{eq:ot6}
\end{eqnarray}
which can be regarded as an effective operator resulting from the
corresponding initial four-quark operator with a single gluon
exchange
\begin{eqnarray}
  O^{(q)}_1\, &=&\,(\bar{q}_ib_i)_{V-A}(\bar{s}_jq_j)_{V-A}\nonumber\\
  &=&(\bar{q}_i\gamma^{\mu}(1-\gamma^{5})b_i)(\bar{s}_j\gamma_{\mu}(1-\gamma^{5})q_j)\label{eq:ot}
\end{eqnarray}
with $q=u,\, c$.

\begin{figure}[h]
\begin{center}
  \includegraphics[scale=0.70]{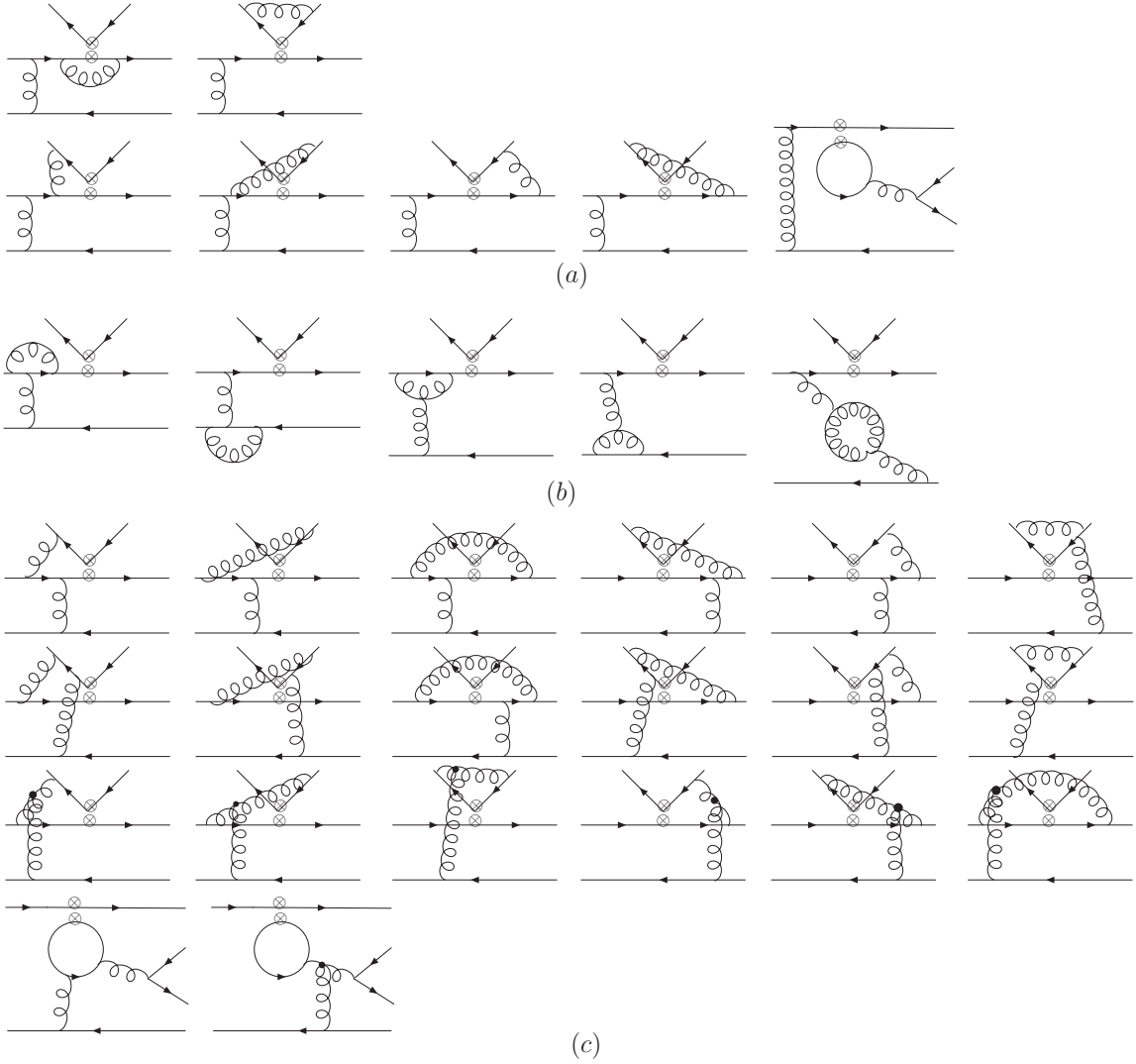}\\
  \caption{
The diagrams in (a) are loop contributions only to the effective
weak vertex
(type I),
 and diagrams in (b) are loop contributions
only to the gluon vertexes (type II). The diagrams in (c) are loop
contributions for both weak and strong vertexes (type III).
  }\label{pic:oneloop}
  \end{center}
\end{figure}

Similar to the procedure of obtaining the four quark operator
effective Hamiltonian from the initial four quark operator
$O_1^{(q)}$ of weak interaction, one should evaluate the six quark
operator effective Hamiltonian from the initial six quark operator
$O_1^{(q)(6)}$ when running the energy scale from $m_{W}$ to the low
energy scale $\mu \sim m_b$. As the first step for finding out the
complete set of independent effective six quark operators, one needs
to evaluate all possible one loop diagrams based on the initial six
quark diagrams (Fig. \ref{pic:4insert}). The possible six quark
diagrams at one loop level are plotted in Fig. \ref{pic:oneloop}.

It is useful to classify those diagrams into three types: type I is
the loop diagrams in which only the effective four quark vertex of
weak interaction receives loop corrections including the penguin
type loops, the single gluon exchanging interaction for six quark
operators remains mediating between one of four external quark lines
of loops and a spectator quark line (see Fig. \ref{pic:oneloop}a);
type II is the loop diagrams where only the single gluon exchanging
vertexes receive loop corrections (see Fig. \ref{pic:oneloop}b); the
remaining loop diagrams are regarded as type III
in which one of the gluon exchanging vertexes touches to the
internal quark/gluon line of loops (see Fig. \ref{pic:oneloop}c).
Note that in Fig. \ref{pic:oneloop}a and Fig. \ref{pic:oneloop}b we
only plot, for an illustration, the six quark diagrams with a gluon
exchanging between one of the four external quark lines of effective
weak vertex and a spectator quark line, while for each of them,
there are actually three additional different diagrams corresponding
to other three choices of external quark lines, they are omitted
just for simplicity.

To evaluate all
the
diagrams is a hard task, as a good approximation, we
shall pay attention to the type I and type II diagrams. The type III
diagrams are in general suppressed at the perturbative region with
energy scale around $m_b$ as they involve more internal quark lines
and contain no large logarithmic enhancements. From the evaluation
of four quark operator effective Hamiltonian, it is known that when
the energy scale runs via the renormalization group evolution from
the high energy scale at $\mu \simeq m_W$ to the low energy scale
around $\mu \sim m_b$, the loop corrections of type I diagrams
should result in the six quark operators with all effective four
quark operators and the corresponding Wilson coefficient functions,
meanwhile the loop corrections of type II diagrams will lead the
strong coupling constant $\alpha_s$ of the gluon exchanging
interaction to run from high energy scale at $m_W$ to the low energy
scale at $\mu$. Thus, when ignoring the type III diagrams, we arrive
at an approximate six quark operator effective Hamiltonian as
follows
\begin{eqnarray}
H_{\rm eff}^{(6)}\, &=&\, \frac{G_F}{\sqrt{2}}\sum_{j=1}^4\{
\sum_{q=u,c}\lambda_q^{s(d)}[C_1(\mu)O_{1q_j}^{(q)(6)}(\mu)
+C_2(\mu)O_{2q_j}^{(q)(6)}(\mu)]\nonumber\\
&& + \sum_{i=3}^{10}\lambda_t^{s(d)} C_i(\mu)O^{(6)}_{i\
q_j}(\mu)\}+h.c.+\dots, \label{eq:hpk6}
\end{eqnarray}
with the CKM factor
$\lambda_q^{s(d)} = V_{qb}V^*_{q s(d)}$.
 The dots represent other possible terms that have been neglected in our
present considerations.
 $O^{(6)}_{i\ q_j}(\mu)$ ($j=1,2,3,4$)
are six quark operators which may effectively be obtained from the
corresponding four quark operators $ O_{i}(\mu)$ (in
Eq.~(\ref{eq:o})) at the scale $\mu$ via the effective gluon
exchanging interactions between one of the external quark lines of
four quark operators and a spectator quark line at the same scale
$\mu$. The general forms and definitions of $O^{(6)}_{i\ q_j}(\mu)$
($j=1,2,3,4$) for the corresponding four quark operators $O_i(\mu)$
are similar to the ones of $O^{(6)}_{q_j}$ ($j=1,2,3,4$) given in
Eq.~(\ref{eq:six}) but with replacing $\alpha_s(m_W)$ by
$\alpha_s(\mu)$ due to QCD corrections of type II diagrams.

Before proceeding, we would like to point out that a complete six
quark operator effective Hamiltonian may involve more effective
operators from the type III diagrams and lead to a non-negligible
contribution to hadronic B meson decays when evaluating the hadronic
matrix elements of six quark operator effective Hamiltonian around
the energy scale $\mu \sim \sqrt{2\Lambda_{QCD} m_b} \sim m_c \sim
1.5$ GeV where the nonperturbative effects may play the role. We
shall keep this in mind and regard the above six quark operator
effective Hamiltonian as an approximate one.

\section{QCD Factorization Based on Effective Six Quark Operators}\label{sec:QCDF}

We shall apply the above effective Hamiltonian with six quark
operators to the nonleptonic two body decays of bottom mesons. The
evaluation of hadronic matrix elements is the most hard task in the
calculations of the decay amplitudes. In this section, we are going
to demonstrate how the factorization approach naturally works for
evaluating the hadronic matrix elements of nonleptonic two body
decays of B meson with six quark operators.

To be explicit, we
here examine
 the hadronic matrix element of $B\to
\pi^0\pi^0$ decay for a typical six quark operator $O^{(6)}_{LL}$
\begin{eqnarray}
O^{(6)}_{LL} & = & \int\!\!\int  \frac{\emph{d}^4k}{(2\pi)^4}\,
  \frac{\emph{d}^4p}{(2\pi)^4}\,e^{-i((x_1-x_2)p+(x_2-x_3)k)}\frac{1}
  {k^2}\ \frac{1}{p^2-m_d^2}  \nonumber \\
  & & [\bar{d}_k(x_2) (p\!\!\!/+m_d)
  \gamma^{\nu}T_{ki}^{a}\gamma^{\mu}(1-\gamma^{5})b_i(x_1)]
  [\bar{d}_j(x_1)\gamma_{\mu}(1-\gamma^{5})d_j(x_1)][\bar{d}_m(x_3)\gamma_{\nu}T_{mn}^{a}
  d_n(x_3)],
\end{eqnarray}
which is actually a part of the six quark operator $O^{(6)}_{4q_2}$
in the effective Hamiltonian. Its hadronic matrix element for $B\to
\pi^0\pi^0$ decay leads to the following most general terms in the
QCD factorization approach

\begin{eqnarray}
  & & M_{LL}^{O}(B\pi\pi) = <\pi^{0} \pi^{0}\mid O^{(6)}_{LL}\mid\bar{B_0}>\,\nonumber\\
& & = \int\!\!\int  \frac{\emph{d}^4k}{(2\pi)^4}\,
  \frac{\emph{d}^4p}{(2\pi)^4}\,e^{-i((x_1-x_2)p+(x_2-x_3)k)}\frac{1}
  {k^2} \  \frac{1}{p^2-m_d^2}  \nonumber \\
  & & <\pi^{0} \pi^{0}\mid [\bar{d}_k(x_2) (p\!\!\!/+m_d)
  \gamma^{\nu}T_{ki}^{a}\gamma^{\mu}(1-\gamma^{5})b_i(x_1)]
  [\bar{d}_j(x_1)\gamma_{\mu}(1-\gamma^{5})d_j(x_1)][\bar{d}_m(x_3)\gamma_{\nu}T_{mn}^{a}
  d_n(x_3)] \mid\bar{B_0}> \nonumber\\
  & &\equiv M_{LL}^{O(1)}+M_{LL}^{O(2)}+M_{LL}^{O(3)}+M_{LL}^{O(4)}\label{eq:loop?},
\end{eqnarray}
with
\begin{eqnarray}
  && M_{LL}^{O(1)}=\int\!\!\int  \frac{\emph{d}^4k}{(2\pi)^4}\,
  \frac{\emph{d}^4p}{(2\pi)^4}\,e^{-i((x_1-x_2)p+(x_2-x_3)k)}
  \frac{1}{k^2(p^2-m_d^2)} T_{ki}^{a}T_{mn}^{a} \nonumber \\
&&\ \,
[(p\!\!\!/+m_d)\gamma^{\nu}\gamma^{\mu}(1-\gamma^{5})]_{\rho\sigma}
  [\gamma_{\mu}(1-\gamma^{5})]_{\alpha\beta}
  [\gamma_{\nu}]_{\gamma\delta} M_{Bim}^{\ \,\sigma\gamma}(x_1,x_3)M_{\pi nk}^{\
\delta\rho}(x_3,x_2)M_{\pi jj}^{\ \beta\alpha}(x_1,x_1),
  \nonumber\\
 & &  M_{LL}^{O(2)}=\int\!\!\int  \frac{\emph{d}^4k}{(2\pi)^4}\,
  \frac{\emph{d}^4p}{(2\pi)^4}\,e^{-i((x_1-x_2)p+(x_2-x_3)k)}
  \frac{1}{k^2(p^2-m_d^2)} T_{ki}^{a}T_{mn}^{a}
 \nonumber \\
&&\ \,
[(p\!\!\!/+m_d)\gamma^{\nu}\gamma^{\mu}(1-\gamma^{5})]_{\rho\sigma}
  [\gamma_{\mu}(1-\gamma^{5})]_{\alpha\beta}
  [\gamma_{\nu}]_{\gamma\delta} M_{Bim}^{\ \,\sigma\gamma}(x_1,x_3)M_{\pi nj}^{\
\delta\alpha}(x_3,x_1)M_{\pi jk}^{\ \beta\rho}(x_1,x_2),
  \nonumber\\
  & & M_{LL}^{O(3)}=\int\!\!\int  \frac{\emph{d}^4k}{(2\pi)^4}\,
  \frac{\emph{d}^4p}{(2\pi)^4}\,e^{-i((x_1-x_2)p+(x_2-x_3)k)}
  \frac{1}{k^2(p^2-m_d^2)} T_{ki}^{a}T_{mn}^{a}
\nonumber \\
&&\ \,
[(p\!\!\!/+m_d)\gamma^{\nu}\gamma^{\mu}(1-\gamma^{5})]_{\rho\sigma}
  [\gamma_{\mu}(1-\gamma^{5})]_{\alpha\beta}
  [\gamma_{\nu}]_{\gamma\delta} M_{Bij}^{\ \,\sigma\alpha}(x_1,x_1)M_{\pi jm}^{\
\beta\gamma}(x_1,x_3)M_{\pi nk}^{\ \delta\rho}(x_3,x_2),
  \nonumber\\
  & & M_{LL}^{O(4)}=\int\!\!\int  \frac{\emph{d}^4k}{(2\pi)^4}\,
  \frac{\emph{d}^4p}{(2\pi)^4}\,e^{-i((x_1-x_2)p+(x_2-x_3)k)}
\frac{1}{k^2(p^2-m_d^2)} T_{ki}^{a}T_{mn}^{a}
\nonumber \\
  & &\ \, [(p\!\!\!/+m_d)\gamma^{\nu}\gamma^{\mu}(1-\gamma^{5})]_{\rho\sigma}
  [\gamma_{\mu}(1-\gamma^{5})]_{\alpha\beta}
  [\gamma_{\nu}]_{\gamma\delta}M_{Bik}^{\ \,\sigma\rho}(x_1,x_2)M_{\pi jm}^{\
\beta\gamma}(x_1,x_3)M_{\pi nj}^{\ \delta\alpha}(x_3,x_1),
\end{eqnarray}
where $M_{X nm}^{\ \, \beta\alpha}(x_i,x_j)\equiv
[M_{X}(x_i,x_j)]_{nm}^{\beta\alpha}$ ($X=B,\pi$) with $n,m$ the
color indices and $\alpha,\beta$ the spinor indices, is the hadronic
matrix element of two quark operators for a single meson $X$. In
light-cone QCD approach, it is found to be \cite{NPB.592.003}
\begin{eqnarray}
  M_{B nm}^{\ \,\beta\alpha}(x_i,x_j)&=&< 0\mid \bar{d}^{\alpha}_{m}(x_j)b^{\beta}_{n}(x_i)\mid \bar{B}^0(P_B)>
  =-\frac{i F_B}{4}\frac{\delta_{mn}}{N_c}\int_0^1 \emph{d}u
  \,e^{-i(u\,P_B^+\,x_j+(P_B-u\,P_B^+)\,x_i)} M_{\text{B}}^{\beta\alpha}(u,P_B),\nonumber\\
  M_{\pi nm}^{\ \beta\alpha}(x_i,x_j)&=&< \pi^{0}(P)\mid \bar{d}^{\alpha}_{m}(x_j) d^{\beta}_{m}(x_i) \mid 0>
  =\frac{iF_{\pi}}{4}\frac{\delta_{mn}}{N_c}\int_0^1 \emph{d}x
  \,e^{-i(x\,P\,x_j+(1-x)P\,x_i)}M_{\pi}^{\beta\alpha}(x,P),
 \label{eq:lcda}
\end{eqnarray}
with $F_{M}$ ($M=B,\pi$) the decay constants. Here
$M_{\text{B}}^{\beta\alpha}(u,P_B)$ and $M_{\pi}^{\beta\alpha}(x,P)$
are the spin structures for the bottom meson and light meson $\pi$
and characterized by the corresponding distribution amplitudes
\begin{eqnarray}
 M_{\text{B}}^{\beta\alpha}(u,P_B)&=&-[m_B+P_B\!\!\slsh\ \ \gamma^5 \phi_{B}(u)]_{\beta\alpha},\nonumber\\
 M_{\pi}^{\beta\alpha}(x,P)&=&[P\,\slsh\,
  \gamma^5 \phi_{\pi}(x)-\mu_{\pi}\gamma^5(\phi_{\pi}^{p}(x)-
  i\sigma_{\mu \nu}n^\mu\slsh v^\nu\slsh \phi^{T}_{\pi}(x)+i\sigma _{\mu \nu}P^{\mu}
\frac{\phi^\sigma(u)}{6} \frac{\partial}{\partial
k_{\bot\nu}})]_{\beta\alpha}\label{eq:daoM},
\end{eqnarray}
with $v=\frac{P}{\sqrt{2}|\overrightarrow{P}|}$, $n=n^+ +n^- -v$ and
$\phi^{T}\equiv \phi^{\sigma\prime}/6$. The light-cone distribution
amplitudes $\phi_M^X(u)$ ($M=B,\pi$, $X=-,p,T$) are given in
\cite{NPB.592.003} up to twist-3. The definition of momentum for
quarks and mesons is explicitly shown in Fig.\ref{pic:definition}.
As a good approximation, both the light quarks and light mesons are
taken to be massless, i.e., $P^2=0$.

\begin{figure}[h]
\begin{center}
  \includegraphics[scale=0.7]{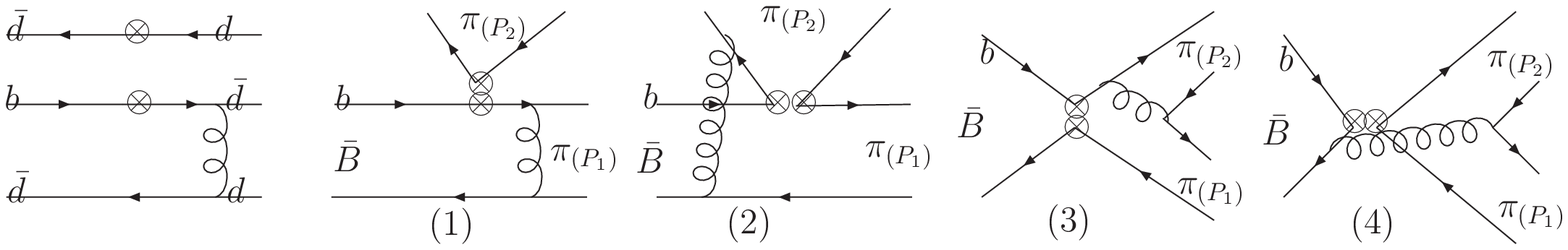}\\
  \caption{
  Different ways of reducing hadronic matrix element of effective six quark operator by QCD factorization approach.
  }\label{pic:Hadronic}
  \end{center}
\end{figure}

It is interesting to note that the four amplitudes $M_{LL}^{O (i)}$
($i=1,2,3,4$) are corresponding to four diagrams (1)-(4) in
Fig.\ref{pic:Hadronic}. The first diagram is known as the
factorizable one, the second is the non-factorizable one and color
suppressed. The third is the factorizable annihilation diagram and
color suppressed, and the fourth is an annihilation diagram and its
matrix element vanishes.

After performing the integration over space-time and momentum, the
above amplitude is simplified to be
\begin{eqnarray}
  && M_{LL}^{O}(B\pi\pi) = <\pi^{0} \pi^{0}\mid  O_{LL}^{(6)}\mid\bar{B_0}> =
  \int_0^1\int_0^1\int_0^1  \emph{d}u\,\emph{d}x\,\emph{d}y\,
  \frac{1}{(u\,P_B^+ -(1-x)P_1)^2}\nonumber\\
  &&[\ \frac{M_{LL}^{(1)}}{(P_{1}-u\,P_B^+)^2-m_d^2}
  +\frac{M_{LL}^{(2)}}{((1-x)P_{1}+y\,P_{2}-u\,P_B^+)^2-m_d^2} \nonumber\\
  && +\frac{M_{LL}^{(3)}}{(x\,P_{1}+P_{2})^2-m_d^2}
  +\frac{M_{LL}^{(4)}}{(x\,P_{1}+(1-y)P_{2}-u\,P_B^+)^2-m_d^2}\ ], \label{eq:ampl}
\end{eqnarray}
with
\begin{eqnarray}
  M_{LL}^{(1)}&=&\frac{C_{F}}{N_c}*F_B\ F_{\pi}^2\text{Tr}[M_{\text{B}}(u,P_B)\gamma_{\nu}
  M_{\pi}(x,P_1)\gamma^{\nu}(P_{1}\!\slsh-u\,P_B^+\!\!\slsh+m_d)\gamma_{\mu}(1-\gamma^{5})] \nonumber \\
  & & \text{Tr}[M_{\pi}(y,P_2)\gamma^{\mu}(1-\gamma^{5})]
  = i\frac{C_{F}}{4N_c}F_B\ F_{\pi}^2 \phi_{B}(u)m_B^3 \mu_{\pi}\phi_{\pi}(y)\phi^{p}_{\pi}(x),\nonumber\\
 M_{LL}^{(2)}&=&\frac{C_{F}}{N_c^2}*F_B\ F_{\pi}^2\text{Tr}[M_{\text{B}}(u,P_B)\gamma_{\nu}
  M_{\pi}(x,P_1)\gamma_{\mu}(1-\gamma^{5})M_{\pi}(y,P_2)\gamma^{\nu}\nonumber\\&&
  ((1-x)P_{1}\!\slsh+y\,P_{2}\!\slsh-u\,P_B^+\!\!\slsh+m_d)\gamma^{\mu}(1-\gamma^{5})]
  \nonumber\\
  &=&i\frac{C_{F}}{4N_c^2}F_B\ F_{\pi}^2 \phi_{B}(u)m_B^3(m_B (u+x+y-2) \phi_{\pi}(x)+\mu_{\pi} (1-x)
  (\phi^{p}_{\pi}(x)-\phi^{T}_{\pi}(x)))\phi_{\pi}(y),\nonumber\\
 M_{LL}^{(3)}&=&\frac{C_{F}}{N_c^2}*F_B\ F_{\pi}^2\text{Tr}[M_{\pi}(x,P_1)\gamma_{\nu}
  M_{\pi}(y,P_2)\gamma^{\nu}(x\,P_{1}\!\slsh+P_{2}\!\slsh+m_d)\gamma_{\mu}(1-\gamma^{5})]\text{Tr}
  [M_{\text{B}}(u,P_B)\gamma^{\mu}(1-\gamma^{5})]
  \nonumber\\
  &=&i\frac{C_{F}}{4N_c^2}F_B\ F_{\pi}^2 \phi_{B}(u)m_B^2(x m_B^2\phi_{\pi}(y) \phi_{\pi}(x)+2
  \mu_{\pi}^2((1+x)\phi^{p}_{\pi}(x)-
  (1-x)\phi^{T}_{\pi}(x))\phi^{p}_{\pi}(y),)\nonumber \\
 M_{LL}^{(4)}&=&0*F_B\ F_{\pi}^2\text{Tr}[M_{\pi}(x,P_1)\gamma_{\nu}M_{\pi}(y,P_2)\gamma_{\mu}(1-\gamma^{5})\nonumber \\
  && M_{\text{B}}(u,P_B)(x\,P_{1}\!\slsh+(1-y)P_{2}\!\slsh-u\,P_B^+\!\!\slsh+m_d)\gamma^{\nu}\gamma^{\mu}(1-\gamma^{5})] = 0,
\end{eqnarray}
where $M_{LL}^{(i)}$ ($i=1,2,3,4$) are obtained by performing the
trace of matrices and determined by the distribution amplitudes.
$C_F=\frac{N_c^2-1}{2N_c}$ is resulted from summing over the color
indices. It can be seen that $M_{LL}^{(1)}$ corresponding to
Fig.\ref{pic:Hadronic}.(1) is color allowed, $M_{LL}^{(2)}$ and
$M_{LL}^{(3)}$ corresponding to Fig.\ref{pic:Hadronic}.(2) and
Fig.\ref{pic:Hadronic}.(3) are color suppressed, while
$M_{LL}^{(4)}$ corresponding to Fig.\ref{pic:Hadronic}.(4) vanishes
as it is not allowed for colorless mesons.

From the above explicit demonstration, it can be seen that the
simple QCD factorization approach becomes a natural tool to evaluate
the hadronic matrix element of effective six quark operators in the
mesonic two body decays. For a given effective six quark operator,
its hadronic matrix element for mesonic two body decays gets four
different combinations in the QCD factorization approach, namely it
consists of four different amplitudes corresponding to four
topologically different diagrams. From the above example, it is
noticed that the amplitude $M_{LL}^{O (1)}$ is a color-allowed
factorizable one in an emission diagram, $M_{LL}^{O (2)}$ is a
color-suppressed non-factorizable one in an emission diagram,
$M_{LL}^{O (3)}$ is a color-suppressed factorizable one in an
annihilation diagram, while $M_{LL}^{O (4)}$ vanishes as it cannot
match to a colorless meson.

\begin{figure}[h]
\begin{center}
  \includegraphics[scale=0.6]{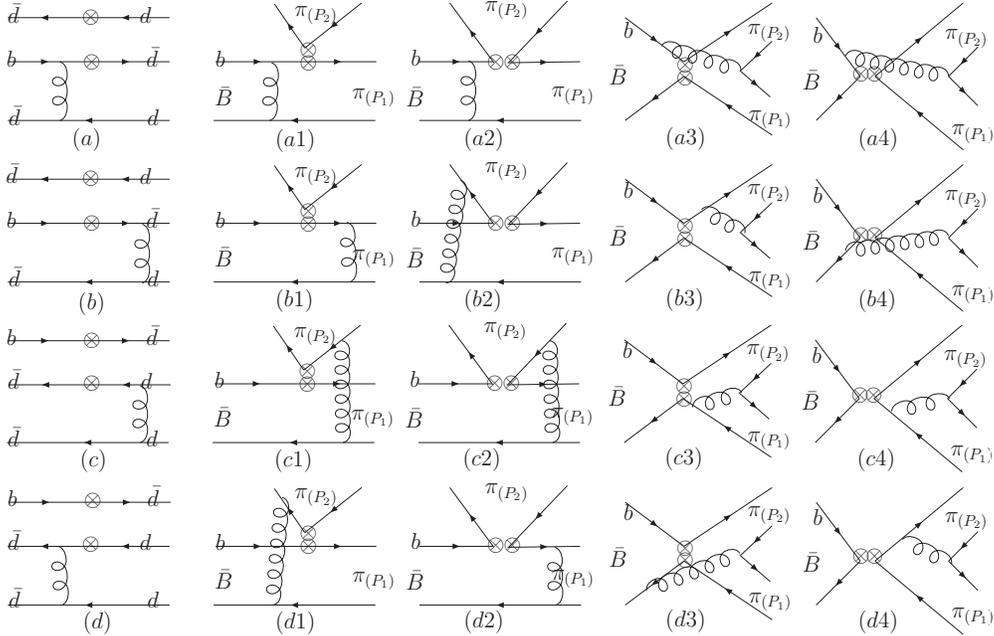}\\
  \caption{Four types of effective six quark diagrams lead to sixteen diagrams
  for hadronic two body decays of heavy meson via QCD factorization.}\label{pic:b-kpi}
  \end{center}
\end{figure}

When generalizing the above analysis to the present framework based
on the approximate six quark operator effective Hamiltonian, there
are in general four types of six quark diagrams corresponding to
four types of effective six quark operators, their hadronic matrix
elements for two body mesonic decays lead to sixteen kinds of
diagrams (see Fig.\ref{pic:b-kpi}.$(ai)$-$(di)$, $i=1,2,3,4)$ as
each of the effective six quark operators leads to four kinds of
amplitudes in the QCD factorization approach.

It is known that the effective four quark vertexes concern three
types of current-current interactions: $(V-A)\times (V-A)$ or
$(LL)$, $(V-A)\times (V+A)$ or $(LR)$, $(S-P)\times (S+P)$ or
$(SP)$, thus each of the diagrams in Fig.\ref{pic:b-kpi} actually
contains three kinds of diagrams corresponding to three types of
current-current interactions. Therefore, there are totally 48 kinds
of hadronic matrix elements involved in the QCD factorization
approach, while it is easy to check that only half of them are
independent with the following relations:
\begin{eqnarray}
\begin{array}{cccccccccccc}
  M^{a1}_{LL}&=&T^F_{LLa};&M^{a2}_{LL}&=&T^F_{LLa}/N_c;&M^{a3}_{LL}&=&A^N_{LLa}/N_c;&M^{a4}_{LL}&=&0;\\
  M^{a1}_{LR}&=&T^F_{LRa};&M^{a2}_{LR}&=&T^F_{SPa}/N_c;&M^{a3}_{LR}&=&A^N_{SPa}/N_c;&M^{a4}_{LR}&=&0;\\
  M^{a1}_{SP}&=&T^F_{SPa};&M^{a2}_{SP}&=&T^F_{LRa}/N_c;&M^{a3}_{SP}&=&A^N_{LRa}/N_c;&M^{a4}_{SP}&=&0;\\
  M^{b1}_{LL}&=&T^F_{LLb};&M^{b2}_{LL}&=&T^N_{LLb}/N_c;&M^{b3}_{LL}&=&A^F_{LLb}/N_c;&M^{b4}_{LL}&=&0;\\
  M^{b1}_{LR}&=&T^F_{LRb};&M^{b2}_{LR}&=&T^N_{SPb}/N_c;&M^{b3}_{LR}&=&A^F_{SPb}/N_c;&M^{b4}_{LR}&=&0;\\
  M^{b1}_{SP}&=&T^F_{LLb};&M^{b2}_{SP}&=&T^N_{LRb}/N_c;&M^{b3}_{SP}&=&A^F_{LRb}/N_c;&M^{b4}_{SP}&=&0;\\
  M^{c1}_{LL}&=&0;&M^{c2}_{LL}&=&T^N_{LLa}/N_c;&M^{c3}_{LL}&=&A^F_{LLa}/N_c;&M^{c4}_{LL}&=&A^F_{LLa};\\
  M^{c1}_{LR}&=&0;&M^{c2}_{LR}&=&T^N_{SPa}/N_c;&M^{c3}_{LR}&=&A^F_{SPa}/N_c;&M^{c4}_{LR}&=&A^F_{LRa};\\
  M^{c1}_{SP}&=&0;&M^{c2}_{SP}&=&T^N_{LRa}/N_c;&M^{c3}_{SP}&=&A^F_{LRa}/N_c;&M^{c4}_{SP}&=&A^F_{SPa};\\
  M^{d1}_{LL}&=&0;&M^{d2}_{LL}&=&T^F_{LLb}/N_c;&M^{d3}_{LL}&=&A^N_{LLb}/N_c;&M^{d4}_{LL}&=&A^F_{LLb};\\
  M^{d1}_{LR}&=&0;&M^{d2}_{LR}&=&T^F_{SPb}/N_c;&M^{d3}_{LL}&=&A^N_{SPb}/N_c;&M^{d4}_{LL}&=&A^F_{LRb};\\
  M^{d1}_{SP}&=&0;&M^{d2}_{SP}&=&T^F_{LRb}/N_c;&M^{d3}_{LL}&=&A^N_{LRb}/N_c;&M^{d4}_{LL}&=&A^F_{SPb}.
\end{array}
\end{eqnarray}
where $T^F_{Xa}$ and  $T^F_{Xb}$ ($X=LL,LR,SP$) represent the
factorizable emission diagram contributions, $T^N_{Xa}$ and
$T^N_{Xb}$ ($X=LL,LR,SP$) are the non-factorizable emission diagram
contributions. $A^F_{Xa}$, $A^F_{Xb}$ and $A^N_{Xa}$, $A^N_{Xb}$
($X=LL,LR,SP$) denote the so-called factorizable and
non-factorizable annihilation diagram contributions respectively.
Their detailed definitions and general formalisms are presented in the
Appendix.

\section{Treatment of Singularities }\label{sec:TOD}

In the evaluation of hadronic matrix elements, there are two kinds
of singularities, one is caused by the infrared divergence of gluon
exchanging interaction, and the other arises from the on-mass shell
divergence of internal quark propagator. As the quark propagator
singularity is a physical-region singularity, one can simply add
$i\epsilon$ to the denominator of quark propagator and apply the
Cutkosky rule~\cite{cutkosky} to avoid such a singularity. It then
allows us to obtain the virtual part of amplitudes as the Cutkosky
rule gives a compact expression for the discontinuity across the cut
arising from a physical-region singularity. In general, a Feynman
diagram will yield an imaginary part for the decay amplitudes when
the virtual particles in the diagram become on mass-shell, and the
resulting diagram can be considered as a genuine physical process.
It is well-known that when applying the Cutkosky rule to deal with a
physical-region singularity of all propagators, the following
formula holds
\begin{eqnarray}
\frac{1}{p^2-m_b^2+i\epsilon}&=&P\biggl[\frac{1}{p^2-m_b^2}
\biggl]-i\pi\delta[p^2-m_b^2],\nonumber\\
\frac{1}{p^2-m_q^2+i\epsilon}&=&P\biggl[\frac{1}{p^2-m_q^2}
\biggl]-i\pi\delta[p^2-m_q^2],\label{quarkd}
\end{eqnarray}
which is known as the principal integration method. Where the first
integration with the notation of capital letter $P$ is the so-called
principal integration.

For the infrared divergence of gluon exchanging interactions, only
adding $i\epsilon$ to the gluon propagator is not enough as such an
infrared divergence is not a physical-region singularity, one cannot
simply apply the Cutkosky rule. To regulate such an infrared
divergence, we may apply the prescription used in the
symmetry-preserving loop regularization\cite{LRC} which allows us to
introduce an intrinsic energy scale without destroying the
non-abelian gauge invariance and translational invariance. The
description of the loop regularization is simple: evaluating the
Feynman integrals to an irreducible integrals, replacing the
integration variable $k^2$ and integration measure
$\int\frac{d^4k}{(2\pi)^4}$ by the regularized ones via \cite{LRC}
\begin{eqnarray}
& & k^2\rightarrow[k^2]_l\equiv k^2-M_l^2, \nonumber \\
& & \int\frac{d^4k}{(2\pi)^4}\rightarrow
\int[\frac{d^4k}{(2\pi)^4}]_l\equiv\lim_{N,M_i^2\to \infty}
\sum_{l=0}^Nc_l^N\int\frac{d^4k}{(2\pi)^4}\label{precedure},
\end{eqnarray}
with conditions
\begin{eqnarray}
& & \lim_{N,M_i^2}\sum_{l=0}^Nc_l^N(M_l^2)^n=0, \quad  c_0^N=1 ~~~~(
i=0,1,\cdots,N~~\mbox{and} ~~n=0,1,\cdots),\label{condition}
\end{eqnarray}
where $c_l^N$ are the coefficients determined by the above
conditions. With a simple form for the regulator masses $M_l = \mu_g
+ l\, M_R$ ($l= 0,1, \cdots$), the coefficients $c_l^N$ is found to
be $ c_l^N = (-1)^l \frac{N!}{(N-l)!\ l!}$, so that
\begin{eqnarray}
k^2~\Rightarrow~k^2-\mu_g^2-l\,M_R^2, \qquad \int
\frac{\emph{d}^4k}{(2\pi)^4}~\Rightarrow~ \lim_{N,M_R\to\infty}
\sum^{N}_{l=0} (-1)^l\frac{N!}{l!(N-l)!}\int
\frac{\emph{d}^4k}{(2\pi)^4},
\end{eqnarray}
which leads the regularized integrals to be independent of the
regulators. Here the energy scale $M_0 = \mu_g$ plays the role
 of infrared cut-off but preserving gauge symmetry and translational
symmetry of original theory.

In the present case, there is no ultraviolet divergence for the
integral over $k$ as it is constrained by the finite momentum of
hadrons, so all the terms with $l\neq 0$ in the summation over $l$
vanish in the limit $M_R\to \infty$. As a consequence, it is
equivalent to add an intrinsic regulator energy scale $\mu_g$ in the
denominator $k^2$ in Eq.~(\ref{eq:loop?}), thus one can use the
usual principal integration method to avoid such a singularity,
i.e.,
\begin{equation}
\frac{1}{k^{2}}~\Rightarrow~\frac{1}{k^{2}-\mu_g^2+i\epsilon}
=P\biggl[\frac{1}{k^{2}-\mu_g^2} \biggl]-i\pi\delta[k^{2}-\mu_g^2].
\end{equation}
With the above considerations, the singularities appearing in the
integrations over $k$ and $p$ can simply be avoided by the following
prescription
\begin{eqnarray}
   \frac{1}{k^2}\frac{1}{(p^2-m^2)} \to
   \frac{1}{(k^2-\mu_g^2+i\epsilon)}\frac{1}{(p^2-m^2+i\epsilon)},
\end{eqnarray}

Note that as the gauge depending team $k_{\mu}k_{\nu}$ can be
transformed to the momentum $p\,\,\slsh$ on exterior line of
spectator quark, they are all on mass shell in our present
consideration (as defined in Fig.\ref{pic:definition} in the
appendix), their contributions equal to zero, thus our results are
gauge independent.

Applying this prescription to the amplitude illustrated in previous
section, we have
\begin{eqnarray}
  && M_{LL}^{O}(B\pi\pi) = <\pi^{0} \pi^{0}\mid  O_{LL}^{(6)}\mid\bar{B_0}> =
  \int_0^1\int_0^1\int_0^1  \emph{d}u\,\emph{d}x\,\emph{d}y\,
  \frac{1}{(u\,P_B^+ -(1-x)P_1)^2-\mu_g^2+i\epsilon}\nonumber\\
  &&[\ \frac{M_{LL}^{(1)}}{(P_{1}-u\,P_B^+)^2-m_{q}^2+i\epsilon}
  +\frac{M_{LL}^{(2)}}{((1-x)P_{1}+y\,P_{2}-u\,P_B^+)^2-m_{q}^2+i\epsilon} \nonumber\\
  && +\frac{M_{LL}^{(3)}}{(x\,P_{1}+P_{2})^2-m_{q}^2+i\epsilon}
  +\frac{M_{LL}^{(4)}}{(x\,P_{1}+(1-y)P_{2}-u\,P_B^+)^2-m_{q}^2+i\epsilon}\ ] .\label{eq:ampl}
\end{eqnarray}

\section{Amplitudes of Charmless Bottom Meson Decays}\label{sec:Amplitude}

With the above considerations and analyses, the QCD factorization
approach enables us to evaluate all the hadronic matrix elements of
nonleptonic two body decays of B meson based on the approximate six
quark operator effective Hamiltonian. The amplitudes of charmless B
meson decays can be expressed as follows:
\begin{eqnarray}
  \label{eq:pptopologyT}
  A(B^0\to \pi^+\pi^-)&=&V_{td}V^*_{tb}[P_T^{\pi\pi}(B)+\frac{2}{3}P_{EW}^{C\pi\pi}(B)+P_E^{\pi\pi}(B)+2P_A^{\pi\pi}(B)+\frac{1}{3}P_{EW}^{A\pi\pi}(B)-\frac{1}{3}A_{EW}^{E\pi\pi}(B)]
  \nonumber\\&&-V_{ud}V^*_{ub}[T^{\pi\pi}(B)+E^{\pi\pi}(B)],\nonumber\\
  A(B^+\to \pi^+\pi^0)&=&\frac{1}{\sqrt{2}}\{V_{td}V^*_{tb}[P_{EW}^{\pi\pi}(B)+P_{EW}^{C\pi\pi}(B)]-V_{ud}V^*_{ub}[T^{\pi\pi}(B)+C^{\pi\pi}(B)]\},\nonumber\\
  A(B^0\to \pi^0\pi^0)&=&\frac{1}{\sqrt{2}}\{-V_{td}V^*_{tb}[P_T^{\pi\pi}(B)-P_{EW}^{\pi\pi}(B)-\frac{1}{3}P_{EW}^{C\pi\pi}(B)+P_E^{\pi\pi}(B)+2P_A^{\pi\pi}(B)
  \nonumber\\&&+\frac{1}{3}P_{EW}^{A\pi\pi}(B)-\frac{1}{3}P_{EW}^{E\pi\pi}(B)]+V_{ud}V^*_{ub}[-C^{\pi\pi}(B)+E^{\pi\pi}(B)]\},
\end{eqnarray}
for $B\to \pi\pi$ decay amplitudes, and
\begin{eqnarray}
  A(B^+\to \pi^+ K^0)&=&-V_{ts}V^*_{tb}[P_T^{\pi K}(B)-\frac{1}{3} P_{EW}^{C\pi K}(B)+P_E^{\pi K}(B)+\frac{2}{3}P_{EW}^{E\pi K}(B)]+V_{us}V^*_{ub}A^{\pi K}(B),\nonumber\\
  A(B^+\to \pi^0 K^+)&=&\frac{1}{\sqrt{2}}\{V_{td}V^*_{tb}[P_T^{\pi K}(B)+P_{EW}^{K\pi}(B)+\frac{2}{3}P_{EW}^{C\pi K}(B)+P_E^{\pi K}(B)+\frac{2}{3}P_{EW}^{E\pi K}(B)]
  \nonumber\\&&-V_{us}V^*_{ub}[T^{\pi K}(B)+C^{K\pi}(B)+A^{\pi K}(B)]\},\nonumber\\
  A(B^0\to \pi^- K^+)&=&V_{td}V^*_{tb}[P_T^{\pi K}(B)+\frac{2}{3}P_{EW}^{C\pi K}(B)+P_E^{\pi K}(B)-\frac{1}{3}P_{EW}^{E\pi K}(B)]-V_{us}V^*_{ub}T^{\pi K}(B),\nonumber\\
  A(B^0\to \pi^0 K^0)&=&-\frac{1}{\sqrt{2}}\{V_{td}V^*_{tb}[P_T^{\pi K}(B)-P_{EW}^{K\pi}(B)-\frac{1}{3}P_{EW}^{C\pi K}(B)+P_E^{\pi K}(B)\nonumber\\&&
  -\frac{1}{3}P_{EW}^{E\pi K}(B)]+V_{us}V^*_{ub}C^{K\pi}(B)\},
\end{eqnarray}
for $B\to \pi K$ decay amplitudes, and
\begin{eqnarray}
  A(B^0\to K^+K^-)&=&-V_{td}V^*_{tb}*[P_A^{K\bar{K}}(B)+P_A^{\bar{K}K}(B)+\frac{2}{3}P_{EW}^{AK\bar{K}}(B)-\frac{1}{3}P_{EW}^{A\bar{K}K}(B)]+V_{ud}V^*_{ub}E^{K\bar{K}}(B),
\nonumber\\
  A(B^+\to K^+\bar{K}^0)&=&-V_{td}V^*_{tb}[P_T^{K\bar{K}}(B)-\frac{1}{3}P_{EW}^{C\pi\pi}(B)+P_E^{K\bar{K}}(B)+\frac{2}{3}P_{EW}^{E\bar{K}K}(B)]+V_{ud}V^*_{ub}A^{K\bar{K}}(B),
\nonumber\\
  A(B^0\to K^0\bar{K}^0)&=&-V_{td}V^*_{tb}[P_T^{K\bar{K}}(B)-\frac{1}{3}P_{EW}^{C\pi\pi}(B)+P_E^{K\bar{K}}(B)+P_A^{K\bar{K}}(B)+P_A^{\bar{K}K}(B)
  \nonumber\\&&-\frac{1}{3}P_{EW}^{AK\bar{K}}(B)-\frac{1}{3}P_{EW}^{A\bar{K}K}(B)-\frac{1}{3}P_{EW}^{EK\bar{K}}(B)],
\end{eqnarray}
for $B\to K K$ decay amplitudes, and
\begin{eqnarray}
  A(B_s^0\to \pi^+\pi^-)&=&-V_{us}V^*_{ub}E^{\pi\pi}(B_s)+V_{ts}V^*_{tb}[2P_{A}^{\pi\pi}(B_s)+\frac{1}{3}P_{EW}^{A\pi\pi}(B_s)],\nonumber\\
  A(B_s^0\to \pi^0\pi^0)&=&\frac{1}{\sqrt{2}}A(B_s^0\to \pi^+\pi^-),\nonumber\\
  A(B_s^0\to \pi^+ K^-)&=&V_{td}V^*_{tb}[P_T^{\bar{K}\pi}(B_s)+\frac{2}{3}P_{EW}^{C\bar{K}\pi}(B_s)+P_{E}^{\bar{K}\pi}(B_s)-\frac{1}{3}P_{EW}^{E\bar{K}\pi}(B_s)]-V_{us}V^*_{ub}T^{\bar{K}\pi}(B_s),\nonumber\\
  A(B_s^0\to \pi^0 K^0)&=&-\frac{1}{\sqrt{2}}\{V_{td}V^*_{tb}[P_T^{\bar{K}\pi}(B_s)-P_{EW}^{\bar{K}\pi}(B_s)-\frac{1}{3}P_{EW}^{C\bar{K}\pi}(B_s)+P_{E}^{\bar{K}\pi}(B_s)-\frac{1}{3}P_{EW}^{E\bar{K}\pi}(B_s)]
  \nonumber\\&&+V_{us}V^*_{ub}C^{\bar{K}\pi}(B_s)\},\nonumber\\
  A(B_s^0\to K^+K^-)&=&-V_{ts}V^*_{tb}[P_T^{\bar{K}K}(B_s)+\frac{2}{3}P_{EW}^{C\bar{K}K}(B_s)+P_E^{\bar{K}K}(B_s)+
  P_{A}^{\bar{K}K}(B_s)+P_{A}^{K\bar{K}}(B_s)\nonumber\\&&+\frac{2}{3}P_{EW}^{AK\bar{K}}(B_s)-\frac{1}{3}P_{EW}^{A\bar{K}K}(B_s)
  -\frac{1}{3}P_{EW}^{E\bar{K}K}(B_s)]+V_{us}V^*_{ub}[T^{\bar{K}K}(B_s)+E^{\bar{K}K}(B_s)],\nonumber\\
  A(B_s^0\to K^0K^0)&=&-V_{ts}V^*_{tb}[P_T^{\bar{K}K}(B_s)-\frac{1}{3}P_{EW}^{C\bar{K}K}(B_s)+P_E^{\bar{K}K}(B_s)+
  P_{A}^{\bar{K}K}(B_s)+P_{A}^{K\bar{K}}(B_s)\nonumber\\&&-\frac{1}{3}P_{EW}^{AK\bar{K}}(B_s)-\frac{1}{3}P_{EW}^{A\bar{K}K}(B_s)
  -\frac{1}{3}P_{EW}^{E\bar{K}K}(B_s)],
\end{eqnarray}
for $B_s\to \pi\pi, \ \pi K,\ KK$ decay amplitudes.
The eleven types of amplitudes
 $T^{M_1M_2}(M)$, $C^{M_1M_2}(M)$,
$P_T^{M_1M_2}(M)$, $P_{EW}^{M_1M_2}(M)$, $A^{M_1M_2}(M)$,
$E^{M_1M_2}(M)$, $P_E^{M_1M_2}(M)$, $P_A^{M_1M_2}(M)$,
$P_{EW}^{CM_1M_2}(M)$, $P_{EW}^{EM_1M_2}(M)$, $P_{EW}^{AM_1M_2}(M)$,
with $M_1M_2 = \pi\pi, \pi K, K\pi, K\bar{K}, \bar{K} K$ are defined
as follows
\begin{eqnarray}
  \label{eq:pptopologyT}
  T^{M_1M_2}(M)&=&4\pi\alpha_s(\mu)\frac{G_F}{\sqrt{2}}\big\{[C_1(\mu)+\frac{1}{N_c}C_2(\mu)]T_{LL}^{FM_1M_2}(M)+\frac{1}{N_c}C_2(\mu)T_{LL}^{NM_1M_2}(M)\big\},\nonumber\\
  C^{M_1M_2}(M)&=&4\pi\alpha_s(\mu)\frac{G_F}{\sqrt{2}}\big\{[C_2(\mu)+\frac{1}{N_c}C_1(\mu)]T_{LL}^{FM_1M_2}(M)+\frac{1}{N_c}C_1(\mu)T_{LL}^{NM_1M_2}(M)\big\},\nonumber\\
  P_T^{M_1M_2}(M)&=&4\pi\alpha_s(\mu)\frac{G_F}{\sqrt{2}}\big\{[C_4(\mu)+\frac{1}{N_c}C_3(\mu)]T_{LL}^{FM_1M_2}(M)+\frac{1}{N_c}C_3(\mu)T_{LL}^{NM_1M_2}(M)
  \nonumber\\&&+[C_6(\mu)+\frac{1}{N_c}C_5(\mu)]T_{SP}^{FM_1M_2}(M)+\frac{1}{N_c}C_5(\mu)T_{LR}^{NM_1M_2}(M)\big\},\nonumber\\
  P_{EW}^{M_1M_2}(M)&=&4\pi\alpha_s(\mu)\frac{G_F}{\sqrt{2}}\frac{3}{2}\big\{[C_9(\mu)+\frac{1}{N_c}C_{10}(\mu)]T_{LL}^{FM_1M_2}(M)+\frac{1}{N_c}C_{10}(\mu)T_{LL}^{NM_1M_2}(M)
  \nonumber\\&&+[C_7(\mu)+\frac{1}{N_c}C_8(\mu)]T_{LR}^{FM_1M_2}(M)+\frac{1}{N_c}C_8(\mu)T_{SP}^{NM_1M_2}(M)\},
  \nonumber\\
  P_{EW}^{CM_1M_2}(M)&=&4\pi\alpha_s(\mu)\frac{G_F}{\sqrt{2}}\frac{3}{2}\big\{[C_{10}(\mu)+\frac{1}{N_c}C_9(\mu)]T_{LL}^{FM_1M_2}(M)+\frac{1}{N_c}C_9(\mu)T_{LL}^{NM_1M_2}(M)
  \nonumber\\&&+[C_8(\mu)+\frac{1}{N_c}C_7(\mu)]T_{SP}^{FM_1M_2}(M)+\frac{1}{N_c}C_7(\mu)T_{LR}^{NM_1M_2}(M)\big\},
\end{eqnarray}
for the so-called emission diagrams, and
\begin{eqnarray}
  \label{eq:pptopologyA}
  A^{M_1M_2}(M)&=&4\pi\alpha_s(\mu)\frac{G_F}{\sqrt{2}}\big\{[C_1(\mu)+\frac{1}{N_c}C_2(\mu)]A_{LL}^{FM_1M_2}(M)+\frac{1}{N_c}C_2(\mu)A_{LL}^{NM_1M_2}(M)\}
  .\nonumber\\
  E^{M_1M_2}(M)&=&4\pi\alpha_s(\mu)\frac{G_F}{\sqrt{2}}\big\{[C_2(\mu)+\frac{1}{N_c}C_1(\mu)]A_{LL}^{FM_1M_2}(M)+\frac{1}{N_c}C_1(\mu)A_{LL}^{NM_1M_2}(M)\big\},\nonumber\\
  P_{E}^{M_1M_2}(M)&=&4\pi\alpha_s(\mu)\frac{G_F}{\sqrt{2}}\big\{[C_4(\mu)+\frac{1}{N_c}C_3(\mu)]A_{LL}^{FM_1M_2}(M)+\frac{1}{N_c}C_3(\mu)A_{LL}^{NM_1M_2}(M)
  \nonumber\\&&+[C_6(\mu)+\frac{1}{N_c}C_5(\mu)]A_{SP}^{FM_1M_2}(M)+\frac{1}{N_c}C_5(\mu)A_{LR}^{NM_1M_2}(M)\}
  ,\nonumber\\
  P_{A}^{M_1M_2}(M)&=&4\pi\alpha_s(\mu)\frac{G_F}{\sqrt{2}}\big\{[C_3(\mu)+\frac{1}{N_c}C_4(\mu)]A_{LL}^{FM_1M_2}(M)+\frac{1}{N_c}C_4(\mu)A_{LL}^{NM_1M_2}(M)
  \nonumber\\&&+[C_5(\mu)+\frac{1}{N_c}C_6(\mu)]A_{LR}^{FM_1M_2}(M)+\frac{1}{N_c}C_6(\mu)A_{SP}^{NM_1M_2}(M)\}
  ,\nonumber\\
  P_{EW}^{AM_1M_2}(M)&=&4\pi\alpha_s(\mu)\frac{G_F}{\sqrt{2}}\frac{3}{2}\big\{[C_9(\mu)+\frac{1}{N_c}C_{10}(\mu)]A_{LL}^{FM_1M_2}(M)+\frac{1}{N_c}C_{10}(\mu)A_{LL}^{NM_1M_2}(M)
  \nonumber\\&&+[C_7(\mu)+\frac{1}{N_c}C_8(\mu)]A_{LR}^{FM_1M_2}(M)+\frac{1}{N_c}C_8(\mu)A_{SP}^{NM_1M_2}(M)\}
  ,\nonumber\\
  P_{EW}^{EM_1M_2}(M)&=&4\pi\alpha_s(\mu)\frac{G_F}{\sqrt{2}}\frac{3}{2}\big\{[C_{10}(\mu)+\frac{1}{N_c}C_9(\mu)]A_{LL}^{FM_1M_2}(M)+\frac{1}{N_c}C_9(\mu)A_{LL}^{NM_1M_2}(M)
  \nonumber\\&&+[C_8(\mu)+\frac{1}{N_c}C_7(\mu)]A_{SP}^{FM_1M_2}(M)+\frac{1}{N_c}C_7(\mu)A_{LR}^{NM_1M_2}(M)\},
\end{eqnarray}
for the so-called annihilation diagrams. Where $T^F_{XA}$,
$T^N_{XA}$, $A^F_{XA}$, $A^N_{XA}$ ($X=LL,LR,SP$, $A=a,b$) arise
from the hadronic matrix elements and their detailed expressions are
given in Appendix. Note that $T_B^{FK\bar{K}}$ and $T_B^{F\bar{K}K}$
are slightly different as the wave functions of $K$ meson and
$\bar{K}$ meson are not equal at high order in the twist expansion.

When redefining the above amplitudes to the widely used diagrammatic
amplitudes in the phenomenological analysis,
\begin{eqnarray}
  &&T=T^{\pi\pi}(B),\  C=C^{\pi\pi}(B),\ E=E^{\pi\pi}(B),\ P=P_T^{\pi\pi}(B)+P_E^{\pi\pi}(B),\ P_A=2P_A^{\pi\pi},\nonumber\\
  &&P_{EW}=P_{EW}^{\pi\pi}(B),\ P_{EW}^C=P_{EW}^{C\pi\pi}(B),\
  P_{EW}^A=P_{EW}^{A\pi\pi}(B),\ P_{EW}^E=P_{EW}^{E\pi\pi}(B),\nonumber\\
  &&T'=T^{\pi K}(B),\ C=C^{K\pi}(B),\ A'=A^{\pi K}(B),\ P'=P_T^{\pi K}(B)+P_E^{\pi K}(B),\ P_A'=2P_A^{K\pi},\nonumber\\
  &&P_{EW}'=P_{EW}^{K\pi}(B),\ P_{EW}'^C=P_{EW}^{C\pi K}(B),\
  P_{EW}'^A=P_{EW}^{AK\pi}(B),\ P_{EW}'^E=P_{EW}^{E\pi K}(B),\\
  &&P''=P_T^{K\bar{K}}(B)+P_E^{K\bar{K}}(B),\
  P_A''=P_A^{K\bar{K}}(B)+P_A^{\bar{K}K}(B),\ P_{EW}''^C=P_{EW}^{CK\bar{K}}(B),\ A''=A^{K\bar{K}}(B),\nonumber\\
  &&P_{EW}''^A=[P_{EW}^{AK\bar{K}}(B) + P_{EW}^{A\bar{K}K}(B)]/2,\
  \tilde{P}_{EW}''^A=[P_{EW}^{AK\bar{K}}(B) - P_{EW}^{A\bar{K}K}(B)]/2,\
  P_{EW}''^E=P_{EW}^{EK\bar{K}}(B).\nonumber
\end{eqnarray}
the decay amplitudes can be reexpressed in terms of the familiar
forms in the diagrammatic decomposition approach
\begin{eqnarray}
  \label{eq:pptopologyT}
  A(B^0\to \pi^+\pi^-)&=&V_{td}V^*_{tb}(P+P_A+\frac{2}{3}P_{EW}^C+\frac{1}{3}P_{EW}^A-\frac{1}{3}P_{EW}^E)-V_{ud}V^*_{ub}(T+E),\nonumber\\
  A(B^+\to \pi^+\pi^0)&=&\frac{1}{\sqrt{2}}[V_{td}V^*_{tb}(P_{EW}+P_{EW}^C)-V_{ud}V^*_{ub}(T+C)],\nonumber\\
  A(B^0\to \pi^0\pi^0)&=&\frac{1}{\sqrt{2}}[V_{td}V^*_{tb}(-P-P_A+P_{EW}+\frac{1}{3}P_{EW}^C+\frac{1}{3}P_{EW}^A-\frac{1}{3}P_{EW}^E)-V_{ud}V^*_{ub}(C-E)],\nonumber\\
  A(B^+\to \pi^+ K^0)&=&V_{ts}V^*_{tb}(P'-\frac{1}{3}P_{EW}'^C+\frac{2}{3}P_{EW}'^E)+V_{us}V^*_{ub}A',\nonumber\\
  A(B^+\to \pi^0 K^+)&=&\frac{1}{\sqrt{2}}[V_{ts}V^*_{tb}(P'+\frac{2}{3}P_{EW}'^C+P_{EW}'+\frac{2}{3}P_{EW}'^E)+V_{us}V^*_{ub}(T'+C'+A')],\nonumber\\
  A(B^0\to \pi^- K^+)&=&V_{ts}V^*_{tb}(P'+\frac{2}{3}P_{EW}'^C-\frac{1}{3}P_{EW}'^E)+V_{us}V^*_{ub}T',\nonumber\\
  A(B^0\to \pi^0
  K^0)&=&\frac{1}{\sqrt{2}}[V_{ts}V^*_{tb}(P'-\frac{1}{3}P_{EW}'^C-P_{EW}'-\frac{1}{3}P_{EW}'^E)+V_{us}V^*_{ub}C'],
  \nonumber \\
  A(B^0\to K^+K^-)&=&V_{td}V^*_{tb}(P_A''+\frac{1}{3}P_{EW}''^A +\tilde{P}_{EW}''^A)+V_{ud}V^*_{ub}E'',\nonumber\\
  A(B^+\to K^+\bar{K}^0)&=&V_{td}V^*_{tb}(P''-\frac{1}{3}P_{EW}''^C+\frac{2}{3}P_{EW}''^E)+V_{ud}V^*_{ub}A'',\nonumber\\
  A(B^0\to K^0\bar{K}^0)&=&V_{td}V^*_{tb}(P''-\frac{1}{3}P_{EW}''^C+P_A''-\frac{2}{3}P_{EW}''^A-\frac{1}{3}P_{EW}''^E).
\end{eqnarray}
It is noticed that there is a slight difference to the usual
diagrammatic decomposition approach with the extra contributions
from the annihilation electro-weak diagrammatic amplitudes
$P_{EW}^{AM_1M_2}$ and $P_{EW}^{EM_1M_2}$, which are actually small
and neglected in the usual diagrammatic decomposition approach.

A similar redefinition can be made for $B_s$ decays,
\begin{eqnarray}
  &&E_s=E^{\pi\pi}(B_s),P_{sA}=2P_A^{\pi\pi}(B_s),P_{sEW}^A=P_{EW}^{A\pi\pi}(B_s),\nonumber\\
  &&T'_s=T^{\bar{K}\pi}(B_s),C'_s=C^{\bar{K}\pi}(B_s),P'_s=P_T^{\bar{K}\pi}(B)+P_E^{\bar{K}\pi}(B_s),P_{sEW}'=P_{EW}^{\bar{K}\pi}(B_s),\nonumber\\
  &&P_{sEW}'^C=P_{EW}^{C\bar{K}\pi}(B_s),P_{sEW}'^E=P_{EW}^{E\bar{K}\pi}(B_s),\nonumber\\
  &&T''_s=T^{\bar{K}K}(B_s),P''_s=P_T^{\bar{K}K}(B_s)+P_E^{\bar{K}K}(B_s),
  P_{sA}''=P_A^{K\bar{K}}(B_s)+P_A^{\bar{K}K}(B_s),P_{sEW}''^C=P_{EW}^{C\bar{K}K}(B),\nonumber\\
  &&E_s''=E^{\bar{K}K}(B_s),P_{sEW}''^A=[P_{EW}^{AK\bar{K}}(B_s) + P_{EW}^{A\bar{K}K}(B_s)]/2,\
  \tilde{P}_{sEW}''^A=[P_{EW}^{AK\bar{K}}(B_s) - P_{EW}^{A\bar{K}K}(B_s)]/2,\nonumber\\
  &&P_{sEW}''^E=P_{EW}^{E\bar{K}K}(B_s).
\end{eqnarray}
Then the
 decay amplitudes can be reexpressed as follows
\begin{eqnarray}
  \label{eq:pptopologyT}
  A(B_s\to \pi^+\pi^-)&=&V_{td}V^*_{tb}(P_{sA}+\frac{1}{3}P_{sEW}^A)-V_{ud}V^*_{ub}E_s,\nonumber\\
  A(B_s\to \pi^0\pi^0)&=&\frac{1}{\sqrt{2}}A(B_s\to \pi^+\pi^-),\nonumber\\
  A(B_s\to \pi^- K^+)&=&V_{ts}V^*_{tb}(P'_s+\frac{2}{3}P_{sEW}'^C-\frac{1}{3}P_{sEW}'^E)+V_{us}V^*_{ub}T'_s,\nonumber\\
  A(B_s\to \pi^0 K^0)&=&-\frac{1}{\sqrt{2}}[V_{ts}V^*_{tb}(P'_s-\frac{1}{3}P_{sEW}'^C-P_{sEW}'-\frac{1}{3}P_{sEW}'^E)+V_{us}V^*_{ub}C'_s],\nonumber\\
  A(B_s\to K^+K^-)&=&V_{td}V^*_{tb}(P_{sA}''+\frac{1}{3}P_{sEW}''^A+\tilde{P}_{sEW}''^A)+V_{ud}V^*_{ub}E''_s,\nonumber\\
  A(B_s\to K^0\bar{K}^0)&=&V_{td}V^*_{tb}(P''_s-\frac{1}{3}P_{sEW}''^C+P_{sA}''-\frac{2}{3}P_{sEW}''^A-\frac{1}{3}P_{sEW}''^E).
\end{eqnarray}

\section{Numerical Calculations}\label{sec:nrcpe}

  We are now in the position to make numerical calculations.

\subsection{Theoretical Input Parameters}

The short distance contributions characterized by the Wilson
coefficient functions for the effective four quark operators were
calculated by several groups at the leading order(LO) and
next-to-leading order(NLO) \cite{hep-ph/9806471}, their values
mainly depend on the choice for the running scale $\mu$. In our
numerical calculations, it is taken to be
\begin{eqnarray}
\mu = \sqrt{2\Lambda_{QCD}m_b} \simeq (1.5\pm 0.1){\rm GeV}.
\end{eqnarray}
The $\alpha_{s}$ value in the six quark operator
effective Hamiltonian is also taken at $\mu = (1.5\pm 0.1)$ GeV.

When considering the NLO Wilson coefficient functions and
$\alpha_{s}$, one needs to include the magnetic penguin-like
operator $O_{8g}$ which is defined as \cite{4qham}
\begin{eqnarray}
O_{8g}\, =\,
\frac{g}{8\pi^2}m_b{\bar{q}}_i\sigma_{\mu\nu}(1+\gamma_5)T_{ij}^aG^{a\mu\nu}b_j\;,
\end{eqnarray}
where $i$, $j$ are the color indices. The magnetic-penguin contribution to
the $B\to\pi K$, $\pi\pi$ decays leads to the modification for the
Wilson coefficients corresponding to the penguin operators,
\begin{eqnarray}
a_{4,6}(\mu)&\to& a_{4,6}(\mu) - \frac{\alpha_s(\mu)}{9\pi}
\frac{2m_B}{\sqrt{|l^2|}}C_{8g}^{\rm eff}(\mu)
\end{eqnarray}
with $C_{8g}^{\rm eff}=C_{8g}+C_5$ and $|l^2| = m_B^2/4$. Where
$a_{4,6}$ are known to be defined as $a_{4,6} =
C_{4,6}+\frac{C_{3,5}}{N_c}$ which appear in the factorizable
diagrams.

For other parameters, we take the following typical values
\begin{eqnarray}
&&m_B = 5.28{\rm GeV},\ m_{\pi^+} = 139.6{\rm MeV},\ m_{\pi^0} = 135{\rm MeV},\ m_b = 4.4{\rm GeV},\ m_c = 1.5{\rm GeV},m_s = 0.1{\rm GeV},\nonumber\\
&&m_u=m_d = 5{\rm MeV},
f_{B} = 216.19{\rm MeV},\ f_{\pi} = 130.1{\rm MeV},\ F_K = 159.8{\rm MeV},
\mu_{\pi} \simeq 1.7{\rm GeV},\nonumber \\& &\mu_K \simeq 1.8{\rm GeV},
 \tau_{B^0} = 1.536 ps,\ \tau_{B^+} = 1.638 ps,\ \lambda =
0.2272,\  A = 0.806,\ \bar{\rho}= 0.195,\ \bar{\eta}=
0.326.\label{def:num}
\end{eqnarray}
Especially, for the infrared energy scale $\mu_g$ introduced in this
paper to regulate the infrared divergence of gluon exchanging
interactions, we take the typical value of $\mu_g$ to be a universal
one around the hadronic bounding energy scale of non-perturbative
QCD
\begin{eqnarray}
\mu_g = (400\pm 50){\rm MeV}.
\end{eqnarray}

To evaluate numerically the hadronic matrix elements of effective
six quark operators based on the QCD factorization, it needs to know
the twist wave functions of mesons. For the wave function of $B$
meson, we take the following form \cite{prd64074004} in our
numerical calculations:
\begin{eqnarray}
\phi_B(x)&=&N_Bx^2(1-x)^2
\exp\left[-\frac{1}{2}\left(\frac{xm_B}{\omega_B}\right)^2 \right]
\;, \label{def:lcda1}
\end{eqnarray}

For the light meson wave functions, it needs to know the twist
distribution amplitudes which contains the twist-2 pion (kaon)
distribution amplitude $\phi_{\pi(K)}$, and the twist-3 ones
$\phi_{\pi(K)}^p$ and $\phi_{\pi(K)}^T$, they are parameterized as
\cite{0508041}:
\begin{eqnarray}
\phi_{\pi(K)}(x) &=& 6x(1-x) (1 + a_1^{\pi(K)} 3 (2x - 1) +
a_2^{\pi(K)}\frac{3}{2}(5(2x-1)^2-1)\nonumber\\
& &+a_4^{\pi(K)}\frac{15}{8}(21(2x-1)^4-14(2x-1)^2+1)) \;,
\\
\phi^p_{\pi(K)}(x) &=&  1 +\left(30\eta_3
-\frac{5}{2}\rho_{\pi(K)}^2\right)
\frac{1}{2}\left(3(2x-1)^2-1\right)
\nonumber\\
&& -\, 3\left\{ \eta_3\omega_3 +
\frac{9}{20}\rho_{\pi(K)}^2(1+6a_2^{\pi(K)}) \right\}
\frac{1}{8}\left(35(2x-1)^4-30(2x-1)^2+3\right) \;,
\\
\phi^T_{\pi(K)}(x) &=& (1-2x)\bigg[ 1 + 6\left(5\eta_3
-\frac{1}{2}\eta_3\omega_3 - \frac{7}{20}
      \rho_{\pi(K)}^2 - \frac{3}{5}\rho_{\pi(K)}^2 a_2^{\pi(K)} \right)\nonumber\\
&&(1-10x+10x^2) \bigg]\;,\label{def:lcda2}
\end{eqnarray}

In our numerical calculations, the shape parameters in the
distribution amplitudes are taken the following typical values:
\begin{eqnarray}
& &\omega_B\, =\,  0.25\; {\rm GeV}\;\;,\omega_{B_s}\, =\,  0.33\; {\rm GeV}\;\;,\;
\eta_3\, =\, 0.015\;\;,\;\omega_3\, =\, -3\;\;,  \nonumber\\
& & a_1^\pi\, =\, 0\;,\;\;\;\;a_1^K\, =\, 0.06\;\;,\;\;\;a_2^K\,=\, 0.10\pm0.10\;,\;\;\;\;
a_2^\pi\, =\, 0.15\pm0.15\;,\;\;\;\; a_4^K\, =a_4^\pi\, =\,0\pm0.10\;.
\label{def:lcda3}
\end{eqnarray}
where the shape parameters for the bottom mesons are taken from
\cite{prd64074004}, and other shape parameters are taken to fit the
data. Since those shape parameters can vary by 100\%, they agree
with the ones in Refs.\cite{0508041, BsPV}, All parameters for the
light mesons are taken at the energy scale 1 GeV \cite{0605112}, run
to our calculation scale. Note that they may vary significantly when
the scale runs to different values.

\subsection{Numerical Results and Discussions}

The numerical results for the CP averaged branching ratios and CP
violations of charmless B meson decays are presented in Table I for
$B\to \pi\pi, \pi K$ decay channels and in Table II for $B\to K
\bar{K}$ channels. In Table III, we give the results for the
branching ratios and CP violations for $B_s \to \pi\pi, \pi K, KK$
decay channels. The LO and NLO are corresponding to the leading
order hadronic matrix elements with the leading order and
next-to-leading order Wilson coefficients(which include the magnetic
penguin-like operator $O_{8g}$). For comparison, the numerical
predictions from the QCDF approach and pQCD approach are also listed
in the Tables. It is seen that most resulting predictions in our
present calculations are in good agreement with experimental data
within the possible uncertainties from both experiments and
theories, while it remains unclear how to understand the puzzles in
the decay channel $B^0\to \pi^0\pi^0$ for its large branching ratio
and possible positive CP violation, and in the decay channel $B\to
\pi^0 K^+$ for the unexpected large positive CP violation.

As shown in\cite{0508041}, adding vertex corrections may improve the
CP violation. The vertex corrections \cite{NPB.606.245} were
proposed to improve the scale dependence of Wilson coefficient
functions. The ref.\cite{NPB.606.245} has considered vertex
corrections that only influence the Wilson coefficients of
factorizable emission amplitudes. Those coefficients are always
combined as $C_{2n-1}+\frac{C_{2n}}{N_c}$ and
$C_{2n}+\frac{C_{2n-1}}{N_c}$ and modified by
\begin{eqnarray}
C_{2n-1}(\mu)+\frac{C_{2n}}{N_c}(\mu) \to
C_{2n-1}(\mu)+\frac{C_{2n}}{N_c}(\mu)
+\frac{\alpha_s(\mu)}{4\pi}C_F\frac{C_{2n}(\mu)}{N_c} V_{2n-1}(M_2)
\;,&&
\nonumber\\
C_{2n}(\mu)+\frac{C_{2n-1}}{N_c}(\mu) \to
C_{2n}(\mu)+\frac{C_{2n-1}}{N_c}(\mu)
+\frac{\alpha_s(\mu)}{4\pi}C_F\frac{C_{2n-1}(\mu)}{N_c} V_{2n}(M_2) \;,&&\nonumber\\
n=1 - 5\;\;&&
\end{eqnarray}
with $M_2$ being the meson emitted from the weak vertex. In the NDR
scheme, $V_i(M)$ are given by \cite{NPB.606.245}
\begin{eqnarray}
V_i(M) &=&\left\{
\begin{array}{ll}
12\ln(\frac{m_b}{\mu})-18+\int_0^1 dx\, \phi_M(x)\, g(x)\;, &
\mbox{\rm for }i=1-4,9,10\;,
\\
-12\ln(\frac{m_b}{\mu})+6-\int_0^1dx\, \phi_M(x)\, g(1-x)\;, &
\mbox{\rm for }i=5,7\;,
\\
-6 +\int_0^1 dx\,\phi_M^{p}(x)\, h(x) \;, &
 \mbox{\rm for }i=6,8\;,
\end{array}
\right.\label{vim}
\end{eqnarray}
$\phi_M(x)$/$\phi_M^p(x)$ is the twist-2/twist-3 meson distribution
amplitudes defined in Eq.~\ref{eq:daoM}. The functions $g(x)/h(x)$
used in the integration are:
\begin{eqnarray}
g(x) &=& 3\left( \frac{1-2x}{1-x}\ln{x} -i\,\pi \right)\nonumber\\
& & +\left[ 2\,{\rm Li}_2(x)-\ln^2 x +\frac{2\ln
x}{1-x}-(3+2i\,\pi)\ln x - (x\leftrightarrow 1-x) \right] \;,
\\
h(x) &=& 2\,{\rm Li}_2(x)-\ln^2 x -(1+2i\,\pi)\ln x -
(x\leftrightarrow 1-x) \;.
\end{eqnarray}

Such a correction does not include the contributions of the first
two diagrams in Fig.2a which are considered as a part of form factor
or meson amplitude. It is interesting to notice that the vertex
corrections do improve the predictions for CP violations and bring
CP violations in the decay channels $B^0\to \pi^0\pi^0$ and $B^+\to
\pi^0 K^+$ to be more close to the experimental data.

To enlarge the branching ratio of $B \to \pi^0 \pi^0$, we shall
examine an interesting case that only two vertexes concerning the
operators $O_1$ and $O_2$ receive additional large nonperturbative
contributions, namely the Wilson coefficients $a_1 =
C_1+\frac{C_2}{N_c}$ and $a_2 = C_2+\frac{C_1}{N_c}$ are modified to
be the effective ones:
\begin{eqnarray}
a_1 \to a_1^{eff} = C_1(\mu)+\frac{C_2}{N_c}(\mu)
+\frac{\alpha_s(\mu)}{4\pi}C_F\frac{C_2(\mu)}{N_c}( V_1(M_2)+V_0)
\;,&&
\nonumber\\
a_2 \to a_2^{eff} = C_2(\mu)+\frac{C_1}{N_c}(\mu)
+\frac{\alpha_s(\mu)}{4\pi}C_F\frac{C_1(\mu)}{N_c} (V_2(M_2)+V_0)
\;,&&
\end{eqnarray}
Taking the value $V_0 = 25$, the resulting branching ratio for $B\to
\pi^0 \pi^0$ becomes consistent with the experimental data.

It is more interesting to consider the possible nonperturbative
effects by taking the effective color number $N_c^{eff}$ in
color-suppressed diagrams. The numerical results with
$N_c^{eff}=1.7$ are presented in Table IV-VI, which provides an
alternative explanation to the puzzle of observed large branching
ratio $B\to \pi^0\pi^0$. For comparison, we also list in Table
\ref{tab:br11}-\ref{tab:br13} the predicted results via the S4
scenario in QCDF \cite{QCDF}.

The method allows us to calculate the relevant transition form
factors at maximal recoil (with NLO Wilson coefficients including
magnetic penguin contribution),
\begin{eqnarray}
  F_0^{B\to\pi}=0.262^{+0.029+0.10}_{-0.024-0.010},\ F_0^{B\to K}=0.322^{+0.034+0.013}_{-0.029-0.011},\ F_0^{B_s\to K}=0.274^{+0.023+0.013}_{-0.013+0.0005},
\label{eq:formfac}
\end{eqnarray}
with input parameters $\mu_g$=400MeV, $\mu$=1.5GeV,
$\mu_{\pi}$=1.7GeV, $\mu_{K}$=1.8GeV.The first error arises from the range for $\mu_g =
350 \sim 450$ MeV, the second error is caused by the running scale
$\mu = 1.4\sim 1.6 $ GeV.

 The resulting form factors agree with the ones obtained from the
light-cone QCD sum rule of heavy quark effective field
theory\cite{0604007}
\begin{eqnarray}
  F_0^{B\to\pi}=0.285^{+0.016}_{-0.015},\  F_0^{B\to K}= 0.345\pm 0.021,\
  F_0^{B_s\to K}=0.296\pm 0.018, \nonumber
\end{eqnarray}
 and from the full QCD sum rule\cite{SR}
\begin{eqnarray}
  F_0^{B\to\pi}=0.258\pm 0.031,\  F_0^{B\to K}= 0.331 \pm 0.041.
  \nonumber
\end{eqnarray}

To know the relative contributions from various diagrams and
hadronic matrix elements of effective six quark operators, we
present in the Table \ref{tab:Ttopo} and Table \ref{tab:Ptopo} the
numerical results for different kinds of topology amplitudes, the
predictions for the strong phases are all relative to the leading
order tree amplitude phase $\delta_T\simeq 1.93$ in $B\to \pi\pi$
decay. It is interesting to see that the amplitudes from the
annihilation diagrams are significant in comparison with the color
suppressed emission diagrams.

The predictions of $S_{\pi^0 K_S}$ in each method are almost the same and obviously
larger than averaged data in PDG. But some new data in \cite{belle,babar} prefer a
larger prediction.

\begin{table}
\begin{center}
\caption{CP averaged branching ratios and CP violations for $B\to
\pi\pi, \pi K$ decay channels. The central values are obtained with
parameters: $\mu_g$=400MeV, $\mu$=1.5GeV, $\mu_{\pi}$=1.7GeV,
$\mu_{K}$=1.8GeV. The first error arises from the range for $\mu_g =
350 \sim 450$ MeV, the second error stems from the running scale
$\mu = 1.4\sim 1.6 $ GeV.}\label{tab:BR1}
\begin{tabular}{l|c|c|cc|ccc}
\hline \hline
\ \ \ \ \ Mode          & Data \cite{HFAG} &QCDF\cite{QCDF}&\multicolumn{2}{c|}{pQCD\cite{0508041}}&\multicolumn{2}{c}{This work} \\
\cline{4-7}
                        &                  &           & LO       & NLO(+MP)        & LO        & NLO(+MP)        \\
\hline
$B^+ \to \pi^+ K^0$     & $ 23.1 \pm 1.0 $ &$19.3^{+1.9+11.3+1.9+13.2}_{-1.9-7.8-2.1-5.6}$       & $17.0$    & $24.1$     & 16.46     &$21.60^{+7.33+4.36}_{-4.86-3.29}$     \\
$B^+\to \pi^0 K^+$      & $ 12.8 \pm 0.6 $ &$11.1^{+1.8+5.8+0.9+6.9}_{-1.7-4.0-1.0-3.0}$       & $10.2$    & $14.0$     & 9.12     & $11.78^{+3.81+2.22}_{-2.53-1.69}$  \\
$B^0 \to \pi^- K^+$     & $ 19.4 \pm 0.6 $ &$16.3^{+2.6+9.6+1.4+11.4}_{-2.3-6.5-1.4-4.8}$       & $14.2$    & $20.5$     & 14.42     &$19.03^{+6.60+3.86}_{-4.39-2.93}$    \\
$B^0 \to \pi^0 K^0 $    & $ 10.0 \pm 0.6 $ &$\,\,7^{+0.7+4.7+0.7+5.4}_{-0.7-3.2-0.7-2.3}$          & $\,\,5.7$ & $\,\,8.7$  & \,\,6.61  & $\,\,8.84^{+3.22+1.89}_{-2.13-1.44}$ \\
\hline
$B^0\to\pi^- \pi^+  $   & $\,\,5.16\pm0.22$&$\,\,8.9^{+4.0+3.6+0.6+1.2}_{-3.4-3.0-1.0-0.8}$        & $\,\,7.0$ & $\,\,6.7$  & \,\,6.63  & $\,\,6.71^{+1.69+0.70}_{-1.24-0.57}$\\
$B^+\to\pi^+\pi^0$      & $\,\,5.7\,\,\pm 0.4$ &$\,\,6.0^{+3.0+2.1+1.0+0.4}_{-2.4-1.8-0.5-0.4}$          & $\,\,3.5$ & $\,\,4.1$  & \,\,4.43  &$\,\,4.69^{+1.03+0.30}_{-0.71-0.26}$  \\
$B^0\to\pi^0\pi^0$      & $\,\,1.31\pm0.21$&$\,\,0.3^{+0.2+0.2+0.3+0.2}_{-0.2-0.1-0.1-0.1}$        & $\,\,\,0.12$& $\,\,0.29$ & \,\,0.11  &$\,\,0.16^{+0.05+0.02}_{-0.05-0.03}$ \\
\hline
$A_{CP}(\pi^+ K^0)$     & $ 0.009 \pm0.025$&$0.009^{+0.002+0.003+0.001+0.006}_{-0.003-0.003-0.001-0.005}$      & $-0.01$   & $-0.01$    &0.016      & $+0.016^{-0.002-0.000}_{+0.003+0.001}$   \\
$A_{CP}(\pi^0 K^+)$     & $ 0.047\pm 0.026$&$0.071^{+0.017+0.020+0.008+0.090}_{-0.018-0.020-0.006-0.097}$      & $-0.08$   & $-0.08$    &-0.093      & $-0.080^{+0.008+0.006}_{-0.004-0.003}$   \\
$A_{CP}(\pi^- K^+)$     & $ -0.095\pm0.013$&$0.043^{+0.011+0.022+0.005+0.087}_{-0.011-0.025-0.006-0.095}$      & $-0.12$   & $-0.10$    &-0.150      & $-0.124^{+0.014+0.008}_{-0.014-0.007}$   \\
$A_{CP}(\pi^0 K^0) $    & $ -0.12\pm 0.11 $&$-0.033^{+0.010+0.013+0.005+0.034}_{-0.008-0.016-0.010-0.033}$     & $-0.02$   & $\,\,0.00$ &-0.006      & $-0.001^{-0.001-0.003}_{+0.000+0.002}$   \\
$S_{\pi^0 K_S}$             & $ 0.58\pm 0.17$\cite{belle,babar} &   \textbf{--}    & $0.70$    & $0.73$     & 0.711      & $ 0.715^{-0.012-0.003}_{+0.002+0.003}$   \\
\hline
$A_{CP}(\pi^- \pi^+)$     & $0.38\pm 0.07$   &$-0.065^{+0.021+0.030+0.001+0.132}_{-0.021-0.028-0.003-0.128}$      & $0.14$    & $0.20$     & 0.178      &  $0.187^{+0.002+0.014}_{-0.001-0.011}$    \\
$A_{CP}(\pi^+\pi^0)$      & $0.04\pm0.05$    &$-0.000^{+0.000+0.000+0.000+0.00}_{-0.000-0.000-0.000-0.000}$    & $0.00$    & $0.00$     &0.000          &   $0.000^{+0.000+0.000}_{-0.000-0.000}$      \\
$A_{CP}(\pi^0\pi^0)$      & $0.36\pm0.32$    &$0.451^{+0.184+0.151+0.043+0.465}_{-0.128-0.138-0.141-0.616}$       & $-0.04$   & $-0.43$    &-0.571      &   $-0.547^{+0.018+0.046}_{-0.025+0.033}$  \\
$S_{\pi\pi}$         & $-0.61\pm0.08$   &    \textbf{--}       & $-0.34$   & $-0.41$    &-0.528      &   $-0.561^{-0.011-0.010}_{+0.011+0.009}$  \\
\hline
 \hline
\end{tabular}
\end{center}

\end{table}

\begin{table}
\begin{center}
\caption{$B\to KK$ modes with the same input parameters as Table
I.}\label{tab:BR2}
\begin{tabular}{l|c|c|c|cc}
\hline \hline
\ \ \ \ \ Mode          & Data \cite{HFAG}  &QCDF\cite{QCDF}&pQCD\cite{0411146}&\multicolumn{2}{c}{This work} \\
\cline{5-6}
                        &                   &               &                       & LO        & NLO(+MP)        \\
\hline
$B^+\to K^+ \bar{K}^0$  & $1.36\pm 0.28$    &$1.36^{+0.45+0.72+0.14+0.91}_{-0.39-0.49-0.15-0.40}$&  1.65       &0.85       &  $1.09^{+0.26+0.18}_{-0.17-0.14}$   \\
$B^0\to K^0 \bar{K}^0$  & $0.96\pm0.20$     &$1.35^{+0.41+0.71+0.13+1.09}_{-0.36-0.48-0.15-0.45}$&  1.75       &0.65       &   $0.84^{+0.22+0.15}_{-0.15-0.12}$   \\
$B^0\to K^+\bar{K}^-  $ & $0.15\pm0.10$     &$0.013^{+0.005+0.008+0.000+0.087}_{-0.005-0.005-0.000-0.011}$& \textbf{--}  &0.07      &   $0.07^{+0.03+0.01}_{-0.03-0.01}$  \\
\hline
$A_{CP}(K^+ \bar{K}^0)$  & $0.12\pm 0.17$    &$-0.163^{+0.047+0.050+0.016+0.113}_{-0.037-0.057-0.017-0.133}$&   \textbf{--}   &0.096       &  $0.078^{+0.013+0.001}_{-0.013-0.001}$   \\
$A_{CP}(K^0 \bar{K}^0)$  & $-0.58\pm0.7$     &$-0.167^{+0.047+0.045+0.015+0.046}_{-0.037-0.051-0.017-0.036}$&   \textbf{--}   &0.000          &   $0.000^{+0.000+0.000}_{-0.000-0.000}$   \\
$A_{CP}(K^+ \bar{K}^-)$  & \textbf{--}       &\textbf{--}    &\textbf{--}    &0.807           &$0.842^{-0.006-0.000}_{+0.042+0.000}$   \\
\hline \hline
\end{tabular}
\end{center}

\end{table}

\begin{table}
\begin{center}
\caption{ $B_s\to \pi\pi, \pi K, KK$ modes with the same input
parameters as Table I.}\label{tab:BR3}
\begin{tabular}{l|c|c|c|cc}
\hline \hline
\ \ \ \ \ Mode          & Data \cite{HFAG}  &QCDF\cite{QCDF}&pQCD\cite{BsPV}&\multicolumn{2}{c}{This work} \\
\cline{5-6}
                        &                   &               &                       & LO        & NLO(+MP)        \\
\hline
$B_s \to \pi^+ \pi^-$    & $0.5\pm 0.5$      &$0.024^{+0.003+0.025+0.000+0.163}_{-0.003-0.012-0.000-0.021}$&$0.57^{+0.16+0.09+0.01}_{-0.13-0.10-0.00}$&0.19       &  $0.23^{+0.01+0.07}_{-0.01-0.05}$   \\
$B_s \to \pi^0 \pi^0$    & \textbf{--}    &$0.012^{+0.001+0.013+0.000+0.082}_{-0.001-0.006-0.000-0.011}$&$0.28^{+0.08+0.04+0.01}_{-0.07-0.05-0.00}$&0.10       & $0.11^{+0.01+0.03}_{-0.01-0.02}$   \\
$B_s \to \pi^+ \bar{K}^-$& $5.0\pm1.25$      &$10.2^{+4.5+3.8+0.7+0.8}_{-3.9-3.2-1.2-0.7}$           &$7.6^{+3.2+0.7+0.5}_{-2.3-0.7-0.5}$&6.96       &   $7.02^{+1.11+0.63}_{-0.91-0.51}$   \\
$B_s \to \pi^0 \bar{K}^0$& \textbf{--}    &$0.49^{+0.28+0.22+0.40+0.33}_{-0.24-0.14-0.14-0.17}$&$0.16^{+0.05+0.10+0.02}_{-0.04-0.05-0.01}$&0.07       & $0.09^{+0.04+0.03}_{-0.03-0.02}$   \\
$B_s \to K^+\bar{K}^-  $ & $24.4\pm4.8$      &$22.7^{+3.5+12.7+2.0+24.1}_{-3.2-8.4-2.0-9.1}$           &$13.6^{+4.2+7.5+0.7}_{-3.2-4.1-0.2}$&13.26      &   $16.68^{+5.37+4.32}_{-3.71-3.24}$  \\
$B_s \to K^0\bar{K}^0  $ & \textbf{--}    &$24.7^{+2.5+13.7+2.6+25.6}_{-2.4-9.2-2.9-9.8}$       &$15.6^{+5.0+8.3+0.0}_{-3.8-4.7-0.0}$&15.25     &  $18.94^{+5.80+4.56}_{-3.96-3.42}$  \\
\hline
$A_{CP}(\pi^+\pi^-    )$& \textbf{--}       & \textbf{--}   &$-0.012^{+0.001+0.012+0.001}_{-0.004-0.012-0.001}$&0.018           &$0.015^{+0.028-0.003}_{-0.020+0.002}$      \\
$A_{CP}(\pi^0\pi^0    )$& \textbf{--}     &   \textbf{--}     &$-0.012^{+0.001+0.012+0.001}_{-0.004-0.012-0.001}$&0.018           & $0.015^{+0.028-0.003}_{-0.020+0.002}$          \\
$A_{CP}(\pi^+\bar{K}^-)$& $0.39\pm 0.17$ &$-0.067^{+0.021+0.031+0.002+0.155}_{-0.022-0.029-0.004-0.152}$&$0.241^{+0.039+0.033+0.023}_{-0.036-0.030-0.012}$&0.182       &  $0.183^{+0.012+0.018}_{-0.009-0.015}$   \\
$A_{CP}(\pi^0\bar{K}^0)$& \textbf{--}     &$0.416^{+0.166+0.143+0.078+0.409}_{-0.120-0.133-0.145-0.510}$&$0.594^{+0.018+0.074+0.022}_{-0.040-0.113-0.035}$ &0.128     &  $-0.054^{+0.014+0.089}_{-0.014-0.081}$   \\
$A_{CP}(K^+\bar{K}^-  )$& \textbf{--}      &$0.040^{+0.010+0.020+0.005+0.104}_{-0.010-0.023-0.005-0.113}$&$-0.23.3^{+0.009+0.049+0.008}_{-0.002-0.044-0.011}$&-0.218           &$-0.185^{+0.014+0.007}_{-0.010-0.009}$ \\
$A_{CP}(K^0\bar{K}^0  )$& \textbf{--}    &$0.009^{+0.002+0.002+0.001+0.002}_{-0.002-0.002-0.001-0.003}$&$0$                                                &0.000           &$0.000^{+0.000+0.000}_{-0.000-0.000}$           \\
\hline \hline
\end{tabular}
\end{center}

\end{table}

\begin{table}[h]
\begin{center}
\caption{The same as Table I but including the vertex
contributions and compare with QCDF S4.}\label{tab:br11}
\begin{tabular}{l|c|c|cc|cccc}
\hline \hline
\ \ \ \ \ Mode          & Data \cite{HFAG}  &QCDF S4\cite{QCDF}&\multicolumn{2}{c|}{pQCD\cite{0508041}}&\multicolumn{4}{c}{This work} \\
\cline{4-9}
                        &                  &           & LO       & NLO+Vertex & LO        & NLO+Vertex     & $a_{1,2}^{eff}$   & $N_c^{eff}$\\
\hline
$B^+ \to \pi^+ K^0$     & $ 23.1 \pm 1.0 $ & $20.3$     & $17.0$    & $24.5^{+13.6\,(+12.9)}_{-\ 8.1\,(-\ 7.8)}$     & 16.45     &$22.06^{+7.39+4.25}_{-4.86-3.21}$  &22.06 &19.50\\
$B^+\to \pi^0 K^+$      & $ 12.8 \pm 0.6 $ & $11.7$     & $10.2$    & $13.9^{+10.0\,(+\ 7.0)}_{-\ 5.6\,(-\ 4.2)}$     & 9.12     &$12.00^{+3.84+2.19}_{-2.54-1.65}$  &11.66 &10.95\\
$B^0 \to \pi^- K^+$     & $ 19.4 \pm 0.6 $ & $18.4$     & $14.2$    & $20.9^{+15.6\,(+11.0)}_{-\ 8.3\,(-\ 6.5)}$     & 14.41     &$19.32^{+6.67+3.84}_{-4.41-2.91}$  &19.62 &18.68 \\
$B^0 \to \pi^0 K^0 $    & $ 10.0 \pm 0.6 $ & $\,\,8.0$  & $\,\,5.7$ & $\,\,9.1^{+\ 5.6\,(+\ 5.1)}_{-\ 3.3\,(-\ 2.9)}$  & \,\,6.61  &$\,\,8.98^{+3.25+1.88}_{-2.14-1.42}$ &\,\,9.70&\,\,8.71\\
\hline
$B^0\to\pi^- \pi^+  $   & $\,\,5.16\pm0.22$& $\,\,5.2$  & $\,\,7.0$ & $\,\,6.5^{+\ 6.7\,(+\ 2.7)}_{-\ 3.8\,(-\ 1.8)}$  & \,\,6.62  &$\,\,7.07^{+1.67+0.71}_{-1.29-0.58}$&5.38&4.89\\
$B^+\to\pi^+\pi^0$      & $\,\,5.7\pm 0.4$ & $\,\,5.1$  & $\,\,3.5$ & $\,\,4.0^{+\ 3.4\,(+\ 1.7)}_{-\ 1.9\,(-\ 1.2)}$  & \,\,4.43  &$\,\,4.27^{+0.96+0.33}_{-0.73-0.29}$&6.98&6.43\\
$B^0\to\pi^0\pi^0$      & $\,\,1.31\pm0.21$& $\,\,0.7$ & $\,\,0.12$& $\,\,0.29^{+0.50\,(+0.13)}_{-0.20\,(-0.08)}$ & \,\,0.11  &$\,\,0.18^{+0.07+0.05}_{-0.04-0.03}$     &1.03&0.98\\
\hline
$A_{CP}(\pi^+ K^0)$     & $ 0.009 \pm0.025$& $0.003$    & $-0.01$   & $-0.01\pm 0.00\,(\pm 0.00)$    &+0.016      & $0.020^{-0.003-0.001}_{+0.003+0.001}$   &0.020&0.018\\
$A_{CP}(\pi^0 K^+)$     & $ 0.047\pm 0.026$& $-0.036$    & $-0.08$   & $-0.01^{+0.03\,(+0.03)}_{-0.05\,(-0.05)}$    &-0.093      &$-0.035^{+0.006+0.004}_{-0.002-0.002}$   &-0.068&-0.0529\\
$A_{CP}(\pi^- K^+)$     & $ -0.095\pm0.013$& $-0.041$    & $-0.12$   & $-0.09^{+0.06\,(+0.04)}_{-0.08\,(-0.06)}$    &-0.150      & $-0.133^{+0.015+0.008}_{-0.011-0.007}$  &-0.117&-0.131\\
$A_{CP}(\pi^0 K^0) $    & $ -0.12\pm 0.11 $& $0.008$    & $-0.02$   & $-0.07^{+0.03\,(+0.01)}_{-0.03\,(-0.01)}$ &-0.006      & $-0.051^{+0.003+0.000}_{-0.002-0.000}$   &0.002&-0.029\\
$S_{\pi^0 K_S}$             & $ 0.58\pm 0.17$\cite{belle,babar}  &   \textbf{--}    & $0.70$& $0.73^{+0.03\,(+0.01)}_{-0.02\,(-0.01)}$     & 0.710      & $ 0.710^{-0.002-0.001}_{-0.002+0.002}$  &0.789 &0.781\\
\hline
$A_{CP}(\pi^- \pi^+)$     & $0.38\pm 0.07$   & $0.103$   & $0.14$    & $0.18^{+0.20\,(+0.07)}_{-0.12\,(-0.06)}$     & 0.178      &  $0.186^{+0.002+0.015}_{-0.002-0.014}$   &0.214 &0.223\\
$A_{CP}(\pi^+\pi^0)$      & $0.04\pm0.05$    & $-0.0002$   & $0.00$    & $0.00\pm 0.00\,(\pm 0.00)$      &0.000          &  $0.000^{+0.000+0.000}_{-0.000-0.0000}$      &0.000 &0.000\\
$A_{CP}(\pi^0\pi^0)$      & $0.36\pm0.32$    & $-0.19$   & $-0.04$   & $0.63^{+0.35\,(+0.09)}_{-0.34\,(-0.15)}$    &-0.571      & $0.470^{+0.010+0.032}_{-0.011-0.018}$ &-0.174 &-0.208\\
$S_{\pi\pi}$              & $-0.61\pm0.08$   &  \textbf{--}  & $-0.34$ & $-0.43^{+1.00\,(+0.05)}_{-0.56\,(-0.05)}$    &-0.528      &  $-0.556^{-0.010-0.009}_{+0.004+0.008}$  &-0.586&-0.479\\
\hline \hline
\end{tabular}
\end{center}

\end{table}

\begin{table}[h]
\begin{center}
\caption{The same as Table II but including the vertex
contributions and compare with QCDF S4.}\label{tab:br12}
\begin{tabular}{l|c|c|c|ccccc}
\hline \hline
\ \ \ \ \ Mode          & Data \cite{HFAG}  &QCDF S4\cite{QCDF}&pQCD\cite{0411146}&\multicolumn{4}{c}{This work} \\
\cline{5-8}
                        &                   &               &                       & LO        & NLO+Vertex      & $a_{1,2}^{eff}$   & $N_c^{eff}$\\
\hline
$B^+\to K^+ \bar{K}^0$  & $1.36\pm 0.28$  &$1.46$&  1.65       &0.85     &$1.13^{+0.26+0.18}_{-0.17-0.14}$  & 1.13&0.85\\
$B^0\to K^0 \bar{K}^0$  & $0.96\pm0.20$  &$1.58$&  1.75        &0.65    &$0.87^{+0.22+0.16}_{-0.14-0.11}$   & 0.87&0.608\\
$B^0\to K^+\bar{K}^-  $ & $0.15\pm0.10$  &$0.070$&  \textbf{--}  &0.07      &  $0.07^{+0.03+0.01}_{-0.10-0.01}$  &0.07&0.29\\
\hline
$A_{CP}(K^+ \bar{K}^0)$  & $0.12\pm 0.17$  &$-0.043$&   \textbf{--}      &0.096    & $0.080^{+0.014+0.002}_{-0.009-0.001}$  &0.080 &0.207\\
$A_{CP}(K^0 \bar{K}^0)$  & $-0.58\pm0.7$   &$-0.115$&   \textbf{--}   &0.000          &  $0.000^{+0.000+0.000}_{-0.000-0.000}$ &0.000&0.000\\
$A_{CP}(K^+ \bar{K}^-)$  &  \textbf{--}        &   \textbf{--}   &   \textbf{--}     &0.807       & $0.842^{-0.005-0.000}_{+0.041+0.000}$ &0.84 &0.78\\
\hline \hline
\end{tabular}
\end{center}

\end{table}

\begin{table}[h]
\begin{center}
\caption{The same as Table III but including the vertex
contributions and compare with QCDF S4.}\label{tab:br13}
\begin{tabular}{l|c|c|c|cccc}
\hline \hline
\ \ \ \ \ Mode          & Data \cite{HFAG}  &QCDF S4\cite{QCDF}&pQCD\cite{BsPV}&\multicolumn{4}{c}{This work} \\
\cline{5-8}
                        &                   &               &                       & LO        & NLO+Vertex      & $a_{1,2}^{eff}$   & $N_c^{eff}$\\
\hline
$B_s \to \pi^+ \pi^-$    & $0.5\pm 0.5$  &$0.155$&$0.57^{+0.16+0.09+0.01}_{-0.13-0.10-0.00}$&0.19       & $0.23^{+0.01+0.07}_{-0.01-0.05}$  &0.23 &0.69\\
$B_s \to \pi^0 \pi^0$    & \textbf{--}    &$0.078$&$0.28^{+0.08+0.04+0.01}_{-0.07-0.05-0.00}$&0.10      & $0.11^{+0.01+0.03}_{-0.01-0.02}$  &0.11 &0.34\\
$B_s \to \pi^+ \bar{K}^-$& $5.0\pm1.25$  &$8.3$       &$7.6^{+3.2+0.7+0.5}_{-2.3-0.7-0.5}$&6.96       & $7.35^{+1.15+0.63}_{-0.94-0.51}$   &5.73&6.58\\
$B_s \to \pi^0 \bar{K}^0$& \textbf{--}    &$0.61$&$0.16^{+0.05+0.10+0.02}_{-0.04-0.05-0.01}$&0.07       & $0.17^{+0.04+0.04}_{-0.03-0.03}$ &0.69 &0.60 \\
$B_s\to K^+\bar{K}^-  $ & $24.4\pm4.8$  &$36.1$       &$13.6^{+4.2+7.5+0.7}_{-3.2-4.1-0.2}$&13.26     &  $16.77^{+5.36+4.23}_{-3.69-3.17}$ &16.97 &15.76\\
$B_s \to K^0\bar{K}^0  $ & \textbf{--}    &$38.3$       &$15.6^{+5.0+8.3+0.0}_{-3.8-4.7-0.0}$&15.25    &  $18.94^{+5.72+4.34}_{-3.89-3.26}$ &18.94 &16.63\\
\hline
$A_{CP}(\pi^+\pi^-    )$& \textbf{--}     &   \textbf{--}     &$-0.012^{+0.001+0.012+0.001}_{-0.004-0.012-0.001}$&0.018           & $0.015^{+0.028-0.003}_{-0.019+0.002}$       &0.015 &0.016 \\
$A_{CP}(\pi^0\pi^0    )$& \textbf{--}     &   \textbf{--}     &$-0.012^{+0.001+0.012+0.001}_{-0.004-0.012-0.001}$&0.018           & $0.015^{+0.028-0.003}_{-0.019+0.002}$       &0.015 &0.016  \\
$A_{CP}(\pi^+\bar{K}^-)$& $0.39\pm 0.17$ &$0.109$&$0.241^{+0.039+0.033+0.023}_{-0.036-0.030-0.012}$&0.182       &  $0.182^{+0.009+0.015}_{-0.002-0.016}$   &0.207&0.171\\
$A_{CP}(\pi^0\bar{K}^0)$& \textbf{--}     &$0.046$&$0.594^{+0.018+0.074+0.022}_{-0.040-0.113-0.035}$ &0.128       &  $0.831^{+0.017+0.017}_{-0.011-0.006}$ &-0.135 &0.057 \\
$A_{CP}(K^+\bar{K}^-  )$& \textbf{--}    &$-0.047$&$-0.23.3^{+0.009+0.049+0.008}_{-0.002-0.044-0.011}$&-0.218           &$-0.194^{+0.014+0.010}_{-0.011-0.010}$     &-0.168    &-0.191  \\
$A_{CP}(K^0\bar{K}^0  )$& \textbf{--}    &$0.006$&$0$                                                &0.000           &$0.000^{+0.000+0.000}_{-0.000-0.000}$        &0.000   &0.000 \\
\hline \hline
\end{tabular}
\end{center}

\end{table}

\begin{table}[h]
\begin{center}
\caption{ Diagrammatic amplitudes relating to the CKM matrix element
$\lambda_u$ with $10^{-7}$ GeV. }\label{tab:Ttopo}
\begin{tabular}{cc|ccccc}
\hline \hline
  \multicolumn{2}{c|}{topology}     &$T$    &$C$                &$A$                &$E$                \\
\hline
            &LO                     & 81.48 & $4.931e^{-0.75i}$ & \textbf{--}       & $7.329e^{-3.07i}$ \\
$\pi\pi$    &NLO(+MP)               & 81.46 & $6.952e^{-0.50i}$ & \textbf{--}       & $7.321e^{-3.07i}$ \\
            &NLO+Vertex             & $83.51e^{0.04i}$ & $13.88e^{-1.60i}$ & \textbf{--}        & $7.321e^{-3.07i}$ \\
\hline
            &LO                     & 100.0 & $6.020e^{-0.75i}$ & $39.86e^{-0.50i}$ &\textbf{--}             \\
$\pi K$     &NLO(+MP)               & 100.0 & $8.558e^{-0.488i}$ & $39.58e^{-0.51i}$ &\textbf{--}           \\
            &NLO+Vertex             & $102.4e^{0.04i}$ & $17.52e^{-1.59i}$ & $39.58e^{-0.50i}$ & \textbf{--}               \\
\hline
            &LO                     & \textbf{--}&\textbf{--}      & $2.267e^{2.30i}$  & $7.169e^{-0.85i}$ \\
$K K$       &NLO(+MP)               & \textbf{--}&\textbf{--}        & $2.045e^{2.31i}$  & $7.088e^{-0.85i}$ \\
            &NLO+Vertex             & \textbf{--}&\textbf{--}     & $2.045e^{2.31i}$  & $7.088e^{-0.85i}$ \\
\hline \hline
\end{tabular}
\end{center}

\end{table}

\begin{table}[h]
\begin{center}
\caption{Diagrammatic amplitudes relating to the CKM matrix element
$\lambda_t$ with $10^{-7}$ GeV.}\label{tab:Ptopo}
\begin{tabular}{cc|ccccccccccccc}
\hline \hline
  \multicolumn{2}{c|}{topology}     &$P_T$              &$P_{EW}$           &$P_{EW}^C$        &$P_A$             &$P_E$             &$P_{EW}^A$         &$P_{EW}^E$         &$P=P_T+P_E$       \\
\hline
            &LO                     & $6.555e^{-3.10i}$ & $1.176e^{-3.13i}$ & $0.101e^{0.50i}$ & $2.210e^{0.06i}$ & $3.137e^{1.51i}$  & $0.021e^{-2.65i}$ & $0.111e^{-0.62i}$& $6.987e^{2.72i}$\\
$\pi\pi$    &NLO(+MP)               & $7.478e^{-3.11i}$ & $1.175e^{-3.13i}$ & $0.076e^{0.69i}$ & $2.442e^{0.06i}$ & $3.339e^{1.51i}$  & $0.019e^{-2.59i}$ & $0.112e^{-0.63i}$& $7.876e^{2.74i}$\\
            &NLO+Vertex             & $7.684e^{-3.07i}$ & $1.201e^{-3.09i}$ & $0.232e^{0.93i}$ & $2.442e^{0.06i}$ & $3.339e^{1.51i}$  & $0.019e^{-2.59i}$ & $0.112e^{-0.63i}$& $7.971e^{2.78i}$\\
\hline
            &LO                     & $8.332e^{-3.10i}$ & $1.490e^{-3.12i}$ & $0.123e^{0.49i}$ & \textbf{--}      & $4.604e^{1.88i}$ & \textbf{--}       & $0.165e^{-0.58i}$ & $10.54e^{2.74i}$  \\
$\pi K$     &NLO(+MP)               & $9.483e^{-3.11i}$ & $1.489e^{-3.12i}$ & $0.092e^{0.68i}$ & \textbf{--}      & $5.069e^{1.91i}$ & \textbf{--}       & $0.150e^{-0.60i}$ & $12.00e^{2.76i}$  \\
            &NLO+Vertex             & $9.731e^{-3.07i}$ & $1.521e^{-3.07i}$ & $0.287e^{0.96i}$ & \textbf{--}      & $5.069e^{1.91i}$ & \textbf{--}       & $0.150e^{-0.60i}$ & $12.11e^{2.79i}$  \\
\hline
            &LO                     & $10.57e^{-3.08i}$ & \textbf{--}       & $0.161e^{0.44i}$ & $1.572e^{0.50i}$ & $3.433e^{1.69i}$ & $0.069e^{-1.26i}$ & $0.053e^{-3.00i}$ & $11.30e^{2.90i}$ \\
$K K$       &NLO(+MP)               & $12.03e^{-3.08i}$ & \textbf{--}       & $0.120e^{0.59i}$ & $1.667e^{0.54i}$ & $3.663e^{1.70i}$ & $0.063e^{-1.31i}$ & $0.052e^{-2.94i}$ & $12.79e^{2.92i}$ \\
            &NLO+Vertex             & $12.35e^{-3.05i}$ & \textbf{--}       & $0.362e^{0.95i}$ & $1.667e^{0.54i}$ & $3.663e^{1.70i}$ & $0.063e^{-1.31i}$ & $0.052e^{-2.94i}$ & $12.99e^{2.95i}$ \\
\hline \hline
\end{tabular}
\end{center}

\end{table}

\begin{table}[h]
\begin{center}
\caption{Diagrammatic amplitudes relating to the CKM matrix element
$\lambda_u$ in $B_s$ decays with $10^{-7}$ GeV. }\label{tab:Ttopos}
\begin{tabular}{cc|ccccc}
\hline \hline
  \multicolumn{2}{c|}{topology}     &$T_s$    &$C_s$                &$A_s$                &$E_s$                \\
\hline
            &LO                     & \textbf{--}& \textbf{--} & \textbf{--}       & $4.027e^{-2.28i}$ \\
$\pi\pi$    &NLO(+MP)               & \textbf{--}& \textbf{--} & \textbf{--}       & $3.980e^{-2.28i}$ \\
            &NLO+Vertex             & \textbf{--}& \textbf{--} & \textbf{--}       & $3.980e^{-3.28i}$ \\
\hline
            &LO                     & $77.75e^{0.08i}$ & $6.221e^{-0.84i}$ & $53.96e^{2.58i}$ &\textbf{--}             \\
$\pi K$     &NLO(+MP)               & $77.74e^{0.08i}$ & $7.876e^{-0.60i}$ & $53.76e^{2.59i}$ &\textbf{--}             \\
            &NLO+Vertex             & $79.63e^{0.08i}$ & $14.96e^{-1.54i}$ & $53.76e^{2.59i}$ &\textbf{--}             \\
\hline
            &LO                     & $95.41e^{0.08i}$ & $8.085e^{-0.84i}$ & $2.705e^{1.77i}$  & $7.313e^{-1.44i}$ \\
$K K$       &NLO(+MP)               & $95.41e^{0.08i}$ & $10.10e^{-0.61i}$ & $2.478e^{1.78i}$  & $7.219e^{-1.43i}$ \\
            &NLO+Vertex             & $97.73e^{0.08i}$ & $19.36e^{-1.50i}$ & $2.478e^{1.78i}$  & $7.219e^{-1.43i}$ \\
\hline \hline
\end{tabular}
\end{center}

\end{table}

\begin{table}[h]
\begin{center}
\caption{Diagrammatic amplitudes relating to the CKM matrix element
$\lambda_t$ in $B_s$ decays with $10^{-7}$ GeV.}\label{tab:Ptopos}
\begin{tabular}{cc|ccccccccccccc}
\hline \hline
  \multicolumn{2}{c|}{topology}     &$P_{T_s}$              &$P_{sEW}$           &$P_{sEW}^C$        &$P_{sA}$             &$P_{sE}$             &$P_{sEW}^A$         &$P_{sEW}^E$         &$P_s=P_{sT}+P_{sE}$       \\
\hline
            &LO                     & \textbf{--}       & \textbf{--}       & \textbf{--}      & $2.365e^{1.00i}$ & \textbf{--}       & $0.026e^{-2.24i}$ & \textbf{--}      & $2.265e^{1.00i}$\\
$\pi\pi$    &NLO(+MP)               & \textbf{--}       & \textbf{--}       & \textbf{--}      & $2.597e^{0.99i}$ & \textbf{--}       & $0.025e^{-2.23i}$ & \textbf{--}      & $2.597e^{0.99i}$\\
            &NLO+Vertex             & \textbf{--}       & \textbf{--}       & \textbf{--}      & $2.597e^{0.99i}$ & \textbf{--}       & $0.025e^{-2.23i}$ & \textbf{--}      & $2.597e^{0.99i}$\\
\hline
            &LO                     & $6.329e^{-3.01i}$ & $1.122e^{-3.06i}$ & $0.108e^{0.79i}$ & \textbf{--}      & $4.105e^{1.12i}$ & \textbf{--}       & $0.044e^{-2.78i}$ & $5.315e^{2.58i}$  \\
$\pi K$     &NLO(+MP)               & $7.215e^{-3.01i}$ & $1.121e^{-3.06i}$ & $0.088e^{1.00i}$ & \textbf{--}      & $4.362e^{1.05i}$ & \textbf{--}       & $0.037e^{-2.25i}$ & $5.745e^{2.62i}$  \\
            &NLO+Vertex             & $7.410e^{-3.02i}$ & $1.145e^{-3.05i}$ & $0.243e^{1.07i}$ & \textbf{--}      & $6.275e^{1.22i}$ & \textbf{--}       & $0.093e^{-1.45i}$ & $7.180e^{2.35i}$  \\
\hline
            &LO                     & $8.231e^{-3.00i}$ & $0.918e^{+3.05i}$ & $0.143e^{0.76i}$ & $1.181e^{1.48i}$ & $5.852e^{1.62i}$ & $0.083e^{-1.47i}$ & $0.043e^{-1.63i}$ & $9.671e^{2.63i}$ \\
$K K$       &NLO(+MP)               & $9.348e^{-3.01i}$ & $0.917e^{+3.05i}$ & $0.118e^{0.95i}$ & $1.320e^{1.50i}$ & $6.238e^{1.62i}$ & $0.078e^{-1.48i}$ & $0.047e^{-1.61i}$ & $10.80e^{2.66i}$ \\
            &NLO+Vertex             & $9.597e^{-3.02i}$ & $0.937e^{+3.04i}$ & $0.307e^{1.04i}$ & $1.320e^{1.50i}$ & $6.238e^{1.62i}$ & $0.078e^{-1.48i}$ & $0.047e^{-1.61i}$ & $10.80e^{2.66i}$ \\
\hline \hline
\end{tabular}
\end{center}

\end{table}

\section{Conclusions}\label{sec:conc}

Based on the approximate six quark operator effective Hamiltonian
derived from perturbative QCD, the QCD factorization approach has
naturally been applied to evaluate the hadronic matrix elements for
charmless two body decays of bottom mesons. The resulting
predictions for the decay amplitudes, branching ratios, and $CP$
asymmetries in $B^0,\ B^+,\ B_s\to \pi\pi,\ \pi K,\ KK$ decay
channels have been found to be consistent with the current
experimental measurements except for a few decay modes.

The puzzles for the observed large branching ratio in $B\to
\pi^0\pi^0$ decay and possible large positive CP violations in $B\to
\pi K^+$ decay need to be further investigated. As we have
emphasized at the beginning that the six quark operator effective
Hamiltonian considered in this paper is an approximate one, and a
large number of six quark diagrams which suppressed at high energy
scales have been ingored, but they may become sizable at low energy
scales. Furthermore, when given the predictions, we have only
considered the uncertainties caused by the choices of running scale
$\mu$ and infrared energy scale $\mu_g$ as their effects are more
significant than others. In general, the theoretical uncertainties
could be much larger when the possible uncertainties for all the
input parameters are included. The masses of light mesons are also
neglected in comparison with the bottom meson masses, i.e,
$m^2_{\pi}/m^2_B \sim 0$ and $m_K^2/m_B^2 \sim 0$.

Nevertheless, it is remarkable that such a simple theoretical
framework based on the approximate six quark operator effective
Hamiltonian from the perturbative QCD and the naive QCD
factorization for the nonperturbative QCD effects can result in a
satisfactory theoretical prediction for the charmless B meson decays
$B,\ B_s \to \pi\pi, \pi K, KK$. It also shows that the singularity
due to the on mass-shell fermion propagator can simply be treated
with the principal integration method by apply the Cutkosky rules,
and the one caused by the gluon exchanging interactions can well be
regulated by the description used in the loop regularization method
with the introduction of an intrinsic energy scale $\mu_g$. In
particular, it is found that such a scale takes a typical value
$\mu_g = (400\pm 50)$ MeV which is around the binding energy of
hadron due to non-perturbative QCD effects.

We would like to point out that although the theoretical framework
discussed above is a much simplified one, it turns out that as the
first order approximation the six quark operator effective
Hamiltonian considered in this paper can be taken as a good starting
point. We have actually examined two interesting cases by
considering teh effective Wilson coefficient functions
$a_{1,2}^{eff}$ and the effective color number $N_c^{eff}$ in the
color suppressed diagrams to bring the prediction for the branching
ratio $B\to \pi^0\pi^0$ be consistent with the experimental data. It
is of interest to calculate high order contributions though it is a
challenging task. On the other hand, the precise measurements of
charmless bottom meson decays, especially the measurements on
CP-violations in $B\to KK$ and $B_s\to \pi\pi, \pi K, KK$ decays,
will provide a useful test for various theoretical frameworks. It is
expected that more and more precise experimental data in the future
super B-factory and LHCb will guide us to arrive at a better
understanding on perturbative and nonperturbative QCD.

\acknowledgments

The authors would like to thank I. Bigi, H.Y. Cheng, A.
Khodjamirian, H.N. Li, G. Ricciardi, C. Sachrajda for useful
discussions and conversations during the KITPC program on Flavor
Physics at Beijing. The author (F.Su) is grateful to M. Beneke for
his kind hospitality. This work was supported in part by the
National Science Foundation of China (NSFC) under the grant
10475105, 10491306, and the key Project of Chinese Academy of
Sciences (CAS).

\appendix*

\section{Calculations of Hadronic Matrix
Elements}\label{sec:CalcHME}

\begin{figure}[h]
\begin{center}
  \includegraphics[scale=0.7]{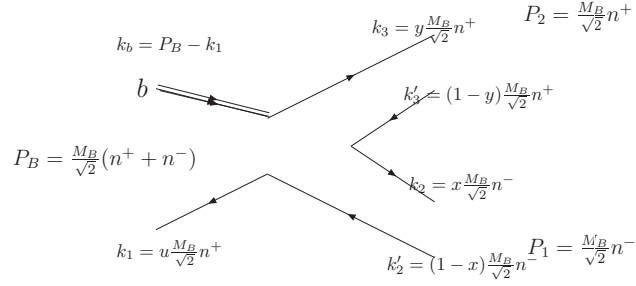}\\
  \caption{Definition of momentum in $B\rightarrow M_1M_2$. The light-cone coordinate is adopted with $(n^+, n^-, k_{\bot})$}\label{pic:definition}
  \end{center}
\end{figure}

In this appendix, we are going to present the explicit expressions
for all the hadronic matrix elements evaluated from the naive QCD
factorization method based on effective six quark operators. To be
specific, we shall first make the following convention for the
momentums of quarks and mesons, which is explicitly shown in Fig.
\ref{pic:definition}

Where we have ignored the light quark mass in external lines and
light meson mass to simplify calculation.

Let us first give the factorizable emission contributions for the
$(V-A)\times (V-A)$ and $(V-A)\times (V+A)$ effective four quark
vertexes, they are simply denoted by $LL$ and $LR$
\begin{eqnarray}
  \label{eq:ppf1}
  T_{LL}^{FM_1M_2}(M)&=&T_{LLa}^{FM_1M_2}(M)+T_{LLb}^{FM_1M_2}(M),\nonumber\\
  T_{LLa}^{FM_1M_2}(M)&=&i \frac{1}{4}\frac{C_{F}}{N_c}\ F_M\ F_{M_1}\
  F_{M_2} \int_0^1\int_0^1\int_0^1  \emph{d}u\,\emph{d}x\,\emph{d}y\,m_B^2\phi_{M}(u)\nonumber\\&&
  \big\{m_B (2 m_b-m_B x) \phi_{M_1}(x)+\mu_{M_1}(2 m_B x-m_b)[\phi^{p}_{M_1}(x)-\phi^{T}_{M_1}(x)]\big\}
  \phi_{M_2}(y)h_{Ta}^F(u,x),\nonumber\\
  T_{LLb}^{FM_1M_2}(M)&=&i \frac{1}{2}  \frac{C_{F}}{N_c}\ F_M\ F_{M_1}\
  F_{M_2} \int_0^1\int_0^1\int_0^1  \emph{d}u\,\emph{d}x\,\emph{d}y\, m_B^3 \mu_{M_1}  \phi_{M}(u)\phi_{M_2}(y) \phi^{p}_{M_1}(x)h_{Tb}^F(u,x),\nonumber\\
  T_{LR}^{FM_1M_2}(M)&=&T_{LLa}^{FM_1M_2}(M)+T_{LLb}^{FM_1M_2}(M),\nonumber\\
  T_{LRa}^{FM_1M_2}(M)&=&- T_{LLa}^{FM_1M_2}(M),\nonumber\\
  T_{LRb}^{FM_1M_2}(M)&=&- T_{LLb}^{FM_1M_2}(M).
\end{eqnarray}
The factorizable emission contributions for the $(S-P)\times (S+P)$
effective four quark vertex are found to be
\begin{eqnarray}
  \label{eq:ppf2}
  T_{SP}^{FM_1M_2}(M)&=&T_{SPa}^{FM_1M_2}(M)+T_{SPb}^{FM_1M_2}(M),\nonumber\\
  T_{SPa}^{FM_1M_2}(M)&=&i \frac{1}{2} \frac{C_{F}}{N_c}\ F_M\ F_{M_1}\
  F_{M_2} \int_0^1\int_0^1\int_0^1  \emph{d}u\,\emph{d}x\,\emph{d}y\, \ m_B\ \mu_{M_2}
  \phi_{M}(u)\nonumber\\&&\big\{m_B(2 m_B-m_b)\phi_{M_1}(x)+\mu_{M_1}[4 m_b- (x+1)m_B] \phi^{p}_{M_1}(x)
  +\mu_{M_1}m_B(1-x)\phi^{T}_{M_1}(x)\big\}\nonumber\\&& \phi^{p}_{M_2}(y)
   h_{Ta}^F(u,x),\nonumber\\
  T_{SPb}^{FM_1M_2}(M)&=&i \frac{1}{2} \frac{C_{F}}{N_c}\ F_M\ F_{M_1}\
  F_{M_2} \int_0^1\int_0^1\int_0^1  \emph{d}u\,\emph{d}x\,\emph{d}y\, \ m_B^2\ \mu_{M_2}\phi_{M}(u)
  \nonumber\\&&[m_B u \phi_{M_1}(x)+2 (1-u)
  \mu_{M_1}\phi^{p}_{M_1}(x)]\phi^{p}_{M_2}(y)h_{Tb}^F(u,x).
\end{eqnarray}
Similarly, we obtain
\begin{eqnarray}
  \label{eq:ppnf1}
  T_{LL}^{NM_1M_2}(M)&=&T_{LLa}^{NM_1M_2}(M)+T_{LLb}^{NM_1M_2}(M),\nonumber\\
  T_{LLa}^{NM_1M_2}(M)&=&-i\frac{1}{4}  \frac{C_{F}}{N_c}\ F_M\ F_{M_1}\ F_{M_2} \int_0^1\int_0^1\int_0^1  \emph{d}u\,\emph{d}x\,\emph{d}y\,
  m_B^3\phi_{M}(u)\nonumber\\&&\big\{ (u-y) m_B\phi_{M_1}(x)
  + (1-x)\mu_{M_1} [\phi^{p}_{M_1}(x)+\phi^{T}_{M_1}(x)]\big\}\phi_{M_2}(y)
  h_{Ta}^N(u,x,y),\nonumber\\
  T_{LLb}^{NM_1M_2}(M)&=&i\frac{1}{4}  \frac{C_{F}}{N_c}\ F_M\ F_1
  \ F_{M_2} \int_0^1\int_0^1\int_0^1  \emph{d}u\,\emph{d}x\,\emph{d}y\,\
  m_B^3\phi_{M}(u)\nonumber\\&&\big\{(u+x+y-2) m_B \phi_{M_1}(x)
  +(1-x)\mu_{M_1}  [\phi^{p}_{M_1}(x)-\phi^{T}_{M_1}(x)]\big\}\phi_{M_2}(y)
  h_{Tb}^N(u,x,y)\nonumber\\
\end{eqnarray}
for non-factorizable emission contributions with the $(V-A)\times
(V-A)$ effective four quark vertex, and
\begin{eqnarray}
  \label{eq:ppnf3}
  T_{LR}^{NM_1M_2}(M)&=&T_{LRa}^{NM_1M_2}(M)+T_{LRb}^{NM_1M_2}(M)\nonumber\\
  T_{LRa}^{NM_1M_2}(M)&=&-i\frac{1}{4} \frac{C_{F}}{N_c}\ F_M\ F_{M_1}\
  F_{M_2} \int_0^1\int_0^1\int_0^1  \emph{d}u\,\emph{d}x\,\emph{d}y\,\ m_B^2\phi_{M}(u)
  \nonumber\\&&\Big\{\mu_{M_2}\ \mu_{M_1}\big\{[(u-x-y+1) \phi^{T}_{M_1}(x)+(u+x-y-1) \phi^{p}_{M_1}(x)]\phi^{p}_{M_2}(y)
  \nonumber\\&&-[(u-x-y+1) \phi^{p}_{M_1}(x)+(u+x-y-1) \phi^{T}_{M_1}(x)]\phi^{T}_{M_2}(y)\}
  \nonumber\\&&+(u-y) m_B\ \mu_{M_2} [\phi^{p}_{M_2}(y)-\phi^{T}_{M_2}(y)]\phi_{M_1}(x)\Big\}h_{Ta}^N(u,x,y)\nonumber\\
  T_{LRb}^{NM_1M_2}(M)&=&i\frac{1}{4} \frac{C_{F}}{N_c}\ F_M\ F_{M_1}\
  F_{M_2} \int_0^1\int_0^1\int_0^1  \emph{d}u\,\emph{d}x\,\emph{d}y\,\ m_B^2\phi_{M}(u)
  \nonumber\\&&\Big\{\mu_{M_2}\ \mu_{M_1}\big\{[(u-x+y)\phi^{T}_{M_1}(x)
  +(u+x+y-2) \phi^{p}_{M_1}(x)] \phi^{p}_{M_2}(y)
  \nonumber\\&&+[(u-x+y)\phi^{p}_{M_1}(x)+(u+x+y-2)\phi^{T}_{M_1}(x)]\phi^{T}_{M_2}(y)\big\}
  \nonumber\\&&+(u+y-1) m_B\ \mu_{M_2}[\phi^{p}_{M_2}(y)+\phi^{T}_{M_2}(y)]\phi_{M_1}(x)\Big\}h_{Tb}^N(u,x,y)
\end{eqnarray}
for non-factorizable emission contributions with the $(V-A)\times
(V+A)$ effective four quark vertex, and
\begin{eqnarray}
  \label{eq:ppnf2}
  T_{SP}^{NM_1M_2}(M)&=&T_{SPa}^{NM_1M_2}(M)+T_{SPb}^{NM_1M_2}(M)\nonumber\\
  T_{SPa}^{NM_1M_2}(M)&=&i \frac{1}{4}\frac{C_{F}}{N_c}\ F_M\ F_{M_1}\
  F_{M_2} \int_0^1\int_0^1\int_0^1  \emph{d}u\,\emph{d}x\,\emph{d}y\,\ m_B^3\phi_{M}(u)\nonumber\\&&
  \big\{(u+x-y-1)m_B  \phi_{M_1}(x)+ (1-x)\mu_{M_1} [\phi^{p}_{M_1}(x)-\phi^{T}_{M_1}(x)]\big\}
  \phi_{M_2}(y)h_{Ta}^N(u,x,y)\nonumber\\
  T_{SPb}^{NM_1M_2}(M)&=&-i \frac{1}{4}\frac{C_{F}}{N_c}\ F_M\ F_{M_1}\
  F_{M_2} \int_0^1\int_0^1\int_0^1  \emph{d}u\,\emph{d}x\,\emph{d}y\,\ m_B^3\phi_{M}(u)\nonumber\\&&
  \big\{(u+y-1)m_B \phi_{M_1}(x)+(1-x)\mu_{M_1} [\phi^{p}_{M_1}(x)+\phi^{T}_{M_1}(x)]\big\}\phi_{M_2}(y)
  h_{Tb}^N(u,x,y)
\end{eqnarray}
for non-factorizable emission contributions with the $(S-P)\times
(S+P)$ effective four quark vertex.

We now present the results from annihilation diagram contributions,
\begin{eqnarray}
  \label{eq:af1}
A_{LL}^{FM_1M_2}(M)&=&A_{LLa}^{FM_1M_2}(M)+A_{LLb}^{FM_1M_2}(M),\nonumber\\
A_{LLa}^{FM_1M_2}(M)&=&i \frac{1}{4} \frac{C_{F}}{N_c}\ F_M\
F_{M_1}\ F_{M_2} \int_0^1\int_0^1\int_0^1
\emph{d}u\,\emph{d}x\,\emph{d}y\,\ m_B^2 \phi_{M}(u)
\nonumber\\&&\big\{(1-y)m_B^2 \phi_{M_2}(y) \phi_{M_1}(x) +2
\mu_{M_2}\ \mu_{M_1} [(2-y) \phi^{p}_{M_2}(y)+y\phi^{T}_{M_2}(y)]
 \phi^{p}_{M_1}(x)\big\}h_{Aa}^F(x,y),\nonumber\\
A_{LLb}^{FM_1M_2}(M)&=&-i \frac{1}{4} \frac{C_{F}}{N_c}\ F_M\
F_{M_1} \ F_{M_2} \int_0^1\int_0^1\int_0^1
\emph{d}u\,\emph{d}x\,\emph{d}y\,\ m_B^2 \phi_{M}(u)
\nonumber\\&&\big\{x m_B^2 \phi_{M_2}(y) \phi_{M_1}(x)+2 \mu_{M_2}\
\mu_{M_1} [(1+x)\phi^{p}_{M_1}(x)-
(1-x)\phi^{T}_{M_1}(x)]\phi^{p}_{M_2}(y)\}
h_{Ab}^F(x,y),\nonumber\\
A_{LR}^{FM_1M_2}(M)&=&A_{LRa}^{FM_1M_2}(M)+A_{LRb}^{FM_1M_2}(M),\nonumber\\
A_{LRa}^{FM_1M_2}(M)&=&A_{LLa}^{FM_1M_2}(M),\nonumber\\
A_{LRb}^{FM_1M_2}(M)&=&A_{LLb}^{FM_1M_2}(M)
\end{eqnarray}
for the factorizable annihilation contributions with the
$(V-A)\times (V-A)$ and $(V-A)\times (V+A)$ effective four quark
vertexes, and
\begin{eqnarray}
  \label{eq:af3}
A_{SP}^{FM_1M_2}(M)&=&A_{SPa}^{FM_1M_2}(M)+A_{SPb}^{FM_1M_2}(M),\nonumber\\
A_{SPa}^{FM_1M_2}(M)&=&i\frac{1}{2} \frac{C_{F}}{N_c}\ F_M\ F_{M_1}\
F_{M_2} \int_0^1\int_0^1\int_0^1  \emph{d}u\,\emph{d}x\,\emph{d}y\,\
m_B^3\phi_{M}(u)
 \nonumber\\&&[(1-y)\mu_{M_2}
[\phi^{p}_{M_2}(y)+\phi^{T}_{M_2}(y)]\phi_{M_1}(x) +2 \mu_{M_1}
\phi_{M_2}(y) \phi^{p}_{M_1}(x)]h_{Aa}^F(x,y),\nonumber\\
A_{SPb}^{FM_1M_2}(M)&=&i\frac{1}{2} \frac{C_{F}}{N_c}\ F_M\ F_{M_1}\
F_{M_2} \int_0^1\int_0^1\int_0^1  \emph{d}u\,\emph{d}x\,\emph{d}y\,\
m_B^3\phi_{M}(u)
 \big\{2 \mu_{M_2}
\phi^{p}_{M_2}(y) \phi_{M_1}(x)\nonumber\\&&+ x\ \mu_{M_1}
\phi_{M_2}(y)[\phi^{p}_{M_1}(x)-\phi^{T}_{M_1}(x)]\big\}h_{Ab}^F(x,y)
\end{eqnarray}
for the factorizable annihilation contributions with the
$(S-P)\times (S+P)$ effective four quark vertex, and
\begin{eqnarray}
  \label{eq:anf1}
A_{LL}^{NM_1M_2}(M)&=&A_{LLa}^{NM_1M_2}(M)+A_{LLb}^{NM_1M_2}(M),\nonumber\\
A_{LLa}^{NM_1M_2}(M)&=&-i\frac{1}{4} \frac{C_{F}}{N_c}\ F_M\
F_{M_1}\ F_{M_2} \int_0^1\int_0^1\int_0^1
\emph{d}u\,\emph{d}x\,\emph{d}y\,\ m_B^2 \phi_{M}(u) \Big\{[m_b+m_B
(u-y)] m_B^2 \phi_{M_2}(y)
\phi_{M_1}(x)\nonumber\\&&+\mu_{M_1}\mu_{M_2} \big\{[-(u-x-y+1)m_B
\phi^{p}_{M_1}(x)+ (-u-x+y+1)m_B\phi^{T}_{M_1}(x)]
 \phi^{T}_{M_2}(y)\big\}\nonumber\\&&+[\big(4 m_b+(u+x-y-1)m_B \big)\phi^{p}_{M_1}(x)+(u-x-y+1)m_B \phi^{T}_{M_1}(x)]
 \phi^{p}_{M_2}(y)\Big\}
 h_{Aa}^N(u,x,y),\nonumber\\
A_{LLb}^{NM_1M_2}(M)&=&i\frac{1}{4} \frac{C_{F}}{N_c}\ F_M\ F_{M_1}\
F_{M_2} \int_0^1\int_0^1\int_0^1  \emph{d}u\,\emph{d}x\,\emph{d}y\,\
m_B^2 \phi_{M}(u) \Big\{x m_B^2 \phi_{M_2}(y)
\phi_{M_1}(x)\nonumber\\&&+\mu_{M_1}\mu_{M_2}\big\{
-[(u+x+y-1)\phi^{p}_{M_1}(x)+(-u+x-y+1)\phi^{T}_{M_1}(x)]
 \phi^{T}_{M_2}(y)\nonumber\\&&+[(-u+x-y+1)\phi^{p}_{M_1}(x)
+(u+x+y-1)\phi^{T}_{M_1}(x)]
 \phi^{p}_{M_2}(y)\big\}\Big\}h_{Ab}^N(u,x,y)
\end{eqnarray}
for the non-factorizable annihilation contributions with the
$(V-A)\times (V-A)$ effective four quark vertex, and
\begin{eqnarray}
  \label{eq:anf3}
A_{LR}^{NM_1M_2}(M)&=&A_{LRa}^{NM_1M_2}(M)+A_{LRb}^{NM_1M_2}(M),\nonumber\\
A_{LRa}^{NM_1M_2}(M)&=&i \frac{1}{4} \frac{C_{F}}{N_c}\ F_M\
F_{M_1}\ F_{M_2} \int_0^1\int_0^1\int_0^1
\emph{d}u\,\emph{d}x\,\emph{d}y\,\ m_B^2\phi_{M}(u)
\nonumber\\&&\big\{\mu_{M_2} [m_b+(y-u)m_B
][\phi^{p}_{M_2}(y)-\phi^{T}_{M_2}(y)]
\phi_{M_1}(x)\nonumber\\&&-\mu_{M_1} [
(1-x)m_B+m_b][\phi^{p}_{M_1}(x)+\phi^{T}_{M_1}(x)]\phi^{p}_{M_2}(y)\big\}
h_{Aa}^N(u,x,y),\nonumber\\
A_{LRa}^{NM_1M_2}(M)&=&-i \frac{1}{4} \frac{C_{F}}{N_c}\ F_M\
F_{M_1}\ F_{M_2} \int_0^1\int_0^1\int_0^1
\emph{d}u\,\emph{d}x\,\emph{d}y\, \ m_B^3\phi_{M}(u)\big\{x\
\mu_{M_1} [\phi^{p}_{M_1}(x)+\phi^{T}_{M_1}(x)]
\phi^{p}_{M_2}(y)\nonumber\\&&-(1-u-y)\mu_{M_2}
[\phi^{p}_{M_2}(y)-\phi^{T}_{M_2}(y)] \phi_{M_1}(x)\}
h_{Ab}^N(u,x,y)
\end{eqnarray}
for the non-factorizable annihilation contributions with the
$(V-A)\times (V-A)$ and $(V-A)\times (V+A)$ effective four quark
vertexes, and
\begin{eqnarray}
  \label{eq:anf2}
A_{SP}^{NM_1M_2}(M)&=&A_{SPa}^{NM_1M_2}(M)+A_{SPb}^{NM_1M_2}(M),\nonumber\\
A_{SPa}^{NM_1M_2}(M)&=&-i \frac{1}{4} \frac{C_{F}}{N_c}\ F_M\
F_{M_1}\  F_{M_2} \int_0^1\int_0^1\int_0^1
\emph{d}u\,\emph{d}x\,\emph{d}y\,\ m_B \phi_{M}(u)
\Big\{[m_b+(x-1)m_B ] m_B^2 \phi^{A}
 _{M_2}(y)\nonumber\\&&+\mu_{M_1}\mu_{M_2} \big\{ \phi_{M_1}(x)[(u-x-y+1)m_B \phi^{p}_{M_1}(x)+
(-u-x+y+1)m_B\phi^{T}_{M_1}(x)]
 \phi^{T}_{M_2}(y)\big\}\nonumber\\&&+\big\{[4 m_b-(-u-x+y+1)m_B ]\phi^{p}_{M_1}(x)-(u-x-y+1)m_B \phi^{T}_{M_1}(x)\big\}
 \phi^{p}_{M_2}(y)\Big\}
 h_{Aa}^N(u,x,y),\nonumber\\
A_{SPb}^{NM_1M_2}(M)&=&i \frac{1}{4} \frac{C_{F}}{N_c}\ F_M\
F_{M_1}\  F_{M_2} \int_0^1\int_0^1\int_0^1
\emph{d}u\,\emph{d}x\,\emph{d}y\,\ m_B^2 \phi_{M}(u) \Big\{(-u-y+1)
m_B^2 \phi_{M_2}(y) \phi_{M_1}(x)\nonumber\\&&+\mu_{M_1}\mu_{M_2}
\big\{[(u+x+y-1)\phi^{p}_{M_1}(x)-(-u+x-y+1)\phi^{T}_{M_1}(x)]
 \phi^{T}_{M_2}(y)\big\}\nonumber\\&&+[(-u+x-y+1)\phi^{p}_{M_1}(x)-(u+x+y-1)\phi^{T}_{M_1}(x)]
 \phi^{p}_{M_2}(y) \Big\} h_{Ab}^N(u,x,y)
\end{eqnarray}
for the non-factorizable annihilation contributions with the
$(S-P)\times (S+P)$ effective four quark vertex.

The functions $h^{Y}_{XA}$ with $(A=a,b)$ from Eqs.~(\ref{eq:ppf1})
to (\ref{eq:anf3}) arise from propagators of gluon and quark, here
$Y=F,N$ denote the factorizable and non-factorizable contributions
respectively, and $X=T,A$ the emission and annihilation diagrams
respectively. They have the following explicit forms:
\begin{eqnarray}
  &&h_{Ta}^F(u,x)=\frac{1}{(-u(1-x)m_B^2-\mu_g^2+i\epsilon)(x m_B^2-m_b^2+i\epsilon)},\nonumber\\
  &&h_{Tb}^F(u,x)=\frac{1}{(-u(1-x)m_B^2-\mu_g^2+i\epsilon)(-u m_B^2-m_{q}^2+i\epsilon)},\nonumber\\
  &&h_{Ta}^N(u,x,y)=\frac{1}{(-u(1-x)m_B^2-\mu_g^2+i\epsilon)((1-x)(1-u-y)m_B^2-m_{q}^2+i\epsilon)},\nonumber\\
  &&h_{Tb}^N(u,x,y)=\frac{1}{(-u(1-x)m_B^2-\mu_g^2+i\epsilon)((1-x)(y-u)m_B^2-m_{q}^2+i\epsilon)},\nonumber\\
  &&h_{Aa}^F(x,y)=\frac{1}{(x(1-y)m_B^2-\mu_g^2+i\epsilon)((1-y)m_B^2-m_{q}^2+i\epsilon)},\nonumber\\
  &&h_{Ab}^F(x,y)=\frac{1}{(x(1-y)m_B^2-\mu_g^2+i\epsilon)(x\,m_B^2-m_{q}^2+i\epsilon)},\nonumber\\
  &&h_{Aa}^N(u,x,y)=\frac{1}{(x(1-y)m_B^2-\mu_g^2+i\epsilon)((y-u)(1-x)m_B^2-m_b^2+i\epsilon)},\nonumber\\
  &&h_{Ab}^N(u,x,y)=\frac{1}{(x(1-y)m_B^2-\mu_g^2+i\epsilon)((1-u-y)x\;m_B^2-m_{q}^2+i\epsilon)}.\label{eq:propN}
\end{eqnarray}

\end{document}